\documentclass[12pt,draftcls,onecolumn]{IEEEtran}

\usepackage{psfrag}
\usepackage{cite}
\usepackage{nccmath}
\usepackage{wasysym,tipa,dblaccnt,accents}
\usepackage{amsmath,amssymb,amsfonts}
\usepackage[T1]{fontenc}
 \usepackage{wasysym,tipa,dblaccnt,accents}
\usepackage{graphicx} % for pdf, bitmapped graphics files
\usepackage{epsfig,subfigure}
\usepackage{stfloats}
\usepackage{url}
\usepackage{mathrsfs}

\input{mathlig}

\DeclareRobustCommand{\scond}{\mathchoice{\muspace\vert\muspace}{\vert}{\vert}{\vert}}
\mathlig{|}{\scond}
\newcommand{\cond}{\,\vert\,}

\newcommand{\Ac}{\mathcal{A}}

\newcommand{\Cc}{\mathcal{C}}
\newcommand{\cc}{\text{\footnotesize $\mathcal{C}$} }

\newcommand{\Ec}{\mathcal{E}}

\newcommand{\Ic}{\mathcal{I}}

\newcommand{\Mc}{\mathcal{M}}

\newcommand{\Sc}{\mathcal{S}}

\newcommand{\Xc}{\mathcal{X}}
\newcommand{\Yc}{\mathcal{Y}}

\newcommand{\aep}{{\mathcal{T}_{\epsilon}^{(n)}}}
\newcommand{\aepvar}{{\mathcal{T}_{\epsilon'}^{(n)}}}

\newcommand{\Mh}{{\hat{M}}}

\newcommand{\Sh}{{\hat{S}}}

\newcommand{\Yh}{{\hat{Y}}}
\newcommand{\Zh}{{\hat{Z}}}

\newcommand{\mh}{{\hat{m}}}
\newcommand{\sh}{{\hat{s}}}

\newcommand{\xh}{{\hat{x}}}
\newcommand{\yh}{{\hat{y}}}

\newcommand{\Ut}{{\tilde{U}}}

\newcommand{\ut}{{\tilde{u}}}

\newcommand{\yt}{{\tilde{y}}}

\def\a{\alpha}
\def\b{\beta}

\def\d{\delta}
\def\e{\epsilon}

\DeclareMathOperator\E{\sf E}
\let\P\relax
\DeclareMathOperator\P{\sf P}
\DeclareMathOperator\C{C}
\DeclareMathOperator\R{R}
\newcommand{\N}{\mathrm{N}}
\newcommand{\U}{\mathrm{Unif}}

\newcommand{\muspace}{\mspace{1mu}}

\newcommand{\sfrac}[2]{\mbox{\small$\displaystyle\frac{#1}{#2}$}}
\newcommand{\half}{\sfrac{1}{2}}

\newtheorem{theorem}{Theorem}
\newtheorem{example}{Example}
\newtheorem{remark}{Remark}
\newtheorem{lemma}{Lemma}
 
\newtheorem{corollary}{Corollary}

\newcommand{\nn}{\nonumber }

\begin{document}

\title{Hybrid Coding: An Interface for Joint Source--Channel Coding and Network Communication}

\author{Paolo~Minero, Sung Hoon Lim, and Young-Han Kim
\thanks{Paolo Minero is with the Department of Electrical Engineering of the University of Notre Dame,  Notre Dame, IN, 46556, USA
(Email: pminero@nd.edu).
Sung Hoon Lim was with the Department of Electrical Engineering, Korea Advanced Institute of
Science and Technology, Daejeon, Korea, and is now with the Samsung Advanced Institute of Technology,
Yongin-si, Gyeonggi-do, Korea
 (Email: sunghlim@kaist.ac.kr).
Young-Han Kim is with the Department of Electrical and Computer Engineering, University of California, San Diego CA, 92093, USA
(Email: yhk@ucsd.edu).}%
 \thanks{This work was supported in part by the National Science Foundation Grant CCF-1117728.}
 \thanks{This work was presented in part at the 2010 Allerton Conference on Communication, Control, and Computing and at the 2011 International Symposium on Information Theory (ISIT 2011).}

\bigskip

\today
}

\maketitle
\allowdisplaybreaks

\begin{abstract}
A new approach to joint source--channel coding is presented in the
 context of communicating correlated sources over multiple access channels.  Similar to the separation architecture, the joint
source--channel coding system architecture in this approach is
modular, whereby the source encoding and channel decoding operations
 are decoupled. However, unlike the separation architecture, the same
codeword is used for both source coding and channel coding, which
allows the resulting hybrid coding scheme to achieve the performance of the
best known joint source--channel coding schemes. Applications of the proposed architecture to
  relay communication are also discussed.
\end{abstract}
\begin{IEEEkeywords}
Analog/digital coding, hybrid coding, joint source--channel coding, network information theory, relay networks.
\end{IEEEkeywords}
 \newpage

%%%%%%%%%%%%%%%%%%%%%%%%%%%%%%%%%%%%%%%%%%
%%%%
%%%%  Introduction
%%%%
%%%%%%%%%%%%%%%%%%%%%%%%%%%%%%%%%%%%%%%%%%

\section{Introduction}
\label{sec:intro}

The fundamental architecture of most of today's communication systems
 is inspired by Shannon's source--channel separation theorem~\cite{Shannon1948, Shannon1959}.
This fundamental theorem states
that a source can be optimally communicated over a point-to-point channel by concatenating an optimal source coder that compresses the source into ``bits'' at the rate of its entropy (or rate--distortion function) with an optimal channel coder that communicates those ``bits''
 reliably over the channel at the rate of its capacity. The appeal of Shannon's separation theorem is twofold.
First, it suggests a simple system architecture in which source coding and channel coding
are separated by a universal digital interface. Second, it guarantees that this
 separation architecture does not incur any asymptotic performance loss.

The optimality of the source--channel separation architecture, however, does not extend to
communication systems with multiple users.
Except for a few special network models in which sources and channels are suitably ``matched''~\cite{Gastpar2003, Song--Yeung--Cai2006, Ramamoorthy2006, agarwal--mitter2010, Jalali--Effros2010, Tian--Diggavi--Shamai2010},
 the problem of lossy communication over a general multiuser network requires the joint optimization of the source coding and channel coding operations. Consequently, there is a vast body of literature on joint source--channel coding schemes for multiple access channels~\cite{Cover--El-Gamal--Salehi1980,deBruyn--prelov1987,Rajesh--etal2008,RajeshSharma2009,Lapidoth--Tinguely2010,Lapidoth--Tinguely2010b,Lim--Minero--Kim2010,Jain-etal2012}, broadcast channels~\cite{Han--Costa1987,Mittal--Phamdo2002,Tuncel2006,Wei--Kramer2008,Kramer--Nair2009,Minero--Kim2009,Kramer--Liang--Shamai2009,Soundararajan-Vishwanath2009,Tian--Diggavi--Shamai2010b,Nayak--Tuncel--Gunduz2010,Gao--Turcel2011b}, interference channels~\cite{Liu--Chen2010,Liu--Chen2011}, and other multiuser channels~\cite{Gunduz--etal2009,Coleman--elal2009,Gunduzetal-2013}.
 % However, matching necessary and sufficient conditions for lossy communication of correlated source over
% these channels are available only in few special cases.
%Mittal--Phamdo2002,Gao--Turcel2011b

This paper takes a new approach to studying the problem of lossy communication of correlated sources over networks. We start by revisiting the problem of transmitting a source over a point-to-point channel, for which we propose a hybrid analog/digital scheme for joint source--channel coding that generalizes both the digital, separate source and channel coding scheme and the analog, uncoded transmission scheme. The proposed hybrid coding scheme employs the architecture depicted in Fig.~\ref{fig:arc} that
 has the following features:
\begin{enumerate}
\item A single code performs both source coding and channel coding.
\item An encoder generates a (digital) codeword from the (analog) source
and selects the channel input as a symbol-by-symbol function of the codeword and the source.
 \item A decoder recovers the (digital) codeword from the (analog) channel output
and selects the source estimate as a symbol-by-symbol function of the codeword and the channel output.
\end{enumerate}

The basic components in this architecture are not new.
 The idea of using a single code for performing both source coding and channel coding appears, for instance, in the celebrated coding scheme by Gelfand and Pinsker~\cite{Gelfand--Pinsker1980a} for channels with state.
The use of symbol-by-symbol functions for the channel input and the source estimate is reminiscent of the Shannon strategy~\cite{Shannon1958a} for channels with states and of the Wyner--Ziv coding scheme~\cite{Gray--Wyner1974} for lossy source coding with side information. Finally, several hybrid analog--digital communication schemes have been  proposed for joint source--channel coding over Gaussian channels; see e.g.,~\cite{Wilson--etal2010}, where the channel input is formed by a combination of digital and analog information.

The main contribution of this paper lies in combining all these known techniques into a unifying framework that can be used to construct coding schemes for various network communication scenarios.
One of the most appealing features of the resulting hybrid coding schemes is that the first-order performance analysis can be done by separately studying the conditions for source coding and for channel coding, exactly as in Shannon's separation theorem. Furthermore, despite its simplicity, hybrid coding yields the best known performance for a general lossy network communication problem (except
 sui generis examples such as \cite{Dueck1981a}). We illustrate the advantages of hybrid coding by focusing on two specific problems:

\begin{figure}[t]
\centering
\small
\psfrag{a}[c]{Analog}
\psfrag{s}[B]{Source}
 \psfrag{m}[B]{Digital}
\psfrag{e1}[c]{Coder}
\psfrag{e3}[c]{Function}
\psfrag{e}[c]{Hybrid encoder}
\psfrag{d}[c]{Hybrid decoder}
\psfrag{d1}[c]{Function}
\psfrag{c}[c]{Channel}
\psfrag{e2}[c]{Decoder}
 \psfrag{mh}[B]{Codeword}
\psfrag{sh}[B]{source estimate}
\includegraphics[scale=0.4]{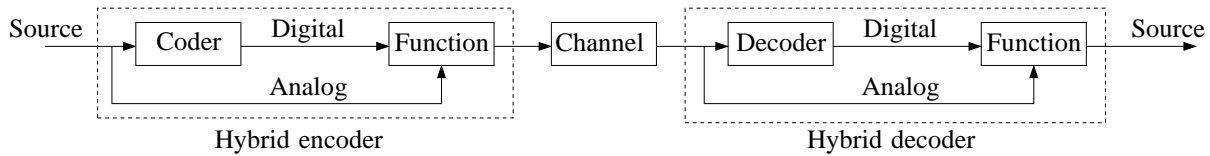}
\caption{A joint source--channel coding system architecture based on hybrid coding.}
\label{fig:arc}
\end{figure}

\begin{enumerate}
\item {\em Joint source--channel coding over multiple access channels}. We construct a joint source--channel coding scheme for lossy communications over multiple access channels  whereby each encoder/decoder in the network operates according to the hybrid coding architecture in Fig.~\ref{fig:arc}. We establish a sufficient condition for lossy communications over multiple access channels that recover and generalize several existing results on joint source--channel coding over this channel model. We also discuss applications of hybrid coding to other channel models such as broadcast channels, interference channels, and channels with feedback.

\item {\em Relay networks.} We apply hybrid coding
beyond joint source--channel coding and propose a new (channel) coding scheme for noisy relay networks.
This coding scheme operates in a similar manner to the noisy network coding scheme proposed in~\cite{Lim--Kim--El-Gamal--Chung2011}, except that
 each relay node uses the hybrid coding interface to transmit a symbol-by-symbol function of the received sequence and its compressed version. This coding scheme unifies both amplify--forward~\cite{schein--gallager2000} and compress--forward~\cite{Cover--El-Gamal1979}, and can strictly outperform both. The potential of the hybrid coding interface for relaying is demonstrated through two specific
 examples---communication over a two--way relay channel~\cite{Rankov--Wittneben2006} and over a  diamond relay network~\cite{schein--gallager2000}.
\end{enumerate}

The remaining of the paper is organized as follows. Section~\ref{sec:ptp} is devoted to the problem of communicating a source over a point-to-point channel. In Section~\ref{sec:networks}, we consider the problem of joint source--channel coding over multiple access channels.  In Section~\ref{sec:relay}, we apply hybrid coding to communication over noisy relay networks. Section~\ref{sec:con} concludes the paper.

Throughout we closely follow the notation in~\cite{El-Gamal--Kim2011}.
In particular, for a discrete random variable $X \sim p(x)$ on an alphabet $\Xc$ and $\e \in (0,1)$, we define the set of $\e$-typical $n$-sequences
 $x^n$ (or the typical set in short)~\cite{Orlitsky--Roche2001} as
$\aep(X) = \{ x^n : | \#\{i : x_i = x\}/n - p(x) | \le \e p(x)
\text{ for all } x \in \Xc \}$. %For a tuple of random variables $(X_1,\ldots,X_k)$,
 %the joint typical set $\aep(X_1,\ldots,X_k)$ is defined as the typical set
%$\aep((X_1,\ldots,X_k))$
%for a single random variable
%$(X_1,\ldots,X_k)$.
We use $\delta(\e) > 0$ to denote a generic function
 of $\e > 0$ that tends to zero as $\e \to 0$. Similarly, we use $\e_n \ge 0$ to denote a generic sequence
in $n$ that tends to zero as $n \to \infty$.

%%%%%%%%%%%%%%%%%%%%%%%%%%%%%%%%%%%%%%%%%%
%%%%
 %%%%  P2P
%%%%
%%%%%%%%%%%%%%%%%%%%%%%%%%%%%%%%%%%%%%%%%%

\section{Point-to-point Channels}
\label{sec:ptp}

\begin{figure}[h!]
\centering
\footnotesize
\psfrag{s}[B]{$S^n$}
\psfrag{m}[B]{$M$}
\psfrag{e1}[c]{\footnotesize Source}
\psfrag{e2}[c]{\footnotesize Channel}
\psfrag{d1}[c]{\footnotesize Source}
 \psfrag{d2}[c]{\footnotesize Channel}
\psfrag{e3}[c]{\footnotesize Source--channel}
\psfrag{d3}[c]{\footnotesize Source--channel}
\psfrag{e}[c]{\footnotesize encoder}
\psfrag{d}[c]{\footnotesize decoder}
\psfrag{e4}[c]{}
 \psfrag{d4}[c]{}
\psfrag{p}[c]{$p(y|x)$}
\psfrag{x}[B]{$X^n$}
\psfrag{y}[B]{$Y^n$}
\psfrag{mh}[B]{$\Mh$}
\psfrag{sh}[B]{$\hat{S}^n$}
\includegraphics[scale=0.35]{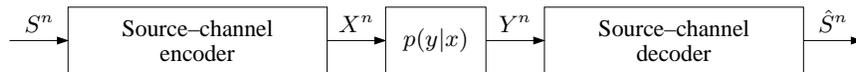}
\caption{Point-to-point communication system.}
 \label{fig:joint}
\end{figure}

Consider the point-to-point communication system depicted in
Fig.~\ref{fig:joint}, where a sender wishes to communicate $n$ symbols of a discrete memoryless source (DMS) $S \sim p(s)$  over the discrete memoryless channel (DMC) $p(y|x)$ in $n$ transmissions so that the receiver can reconstruct the source symbols with a prescribed distortion $D$.

%%%%%%%

An $(|\Sc|^n, n)$ joint source--channel code consists of
\begin{itemize}
\item an encoder that assigns a sequence
  $x^n(s^n)\in \Xc^n$ to each sequence $s^n\in \Sc^n$ and
 \item a decoder that assigns an estimate $\hat{s}^n \in
  \hat\Sc^n$ to each sequence $y^n\in \Yc^n$.
\end{itemize}
Let $d(s,\sh)$  be a nonnegative distortion measure that quantifies the cost of representing a symbol $s$ by a symbol $\sh$.  A distortion $D$ is said to be achievable
 for communication of the DMS $S$ over the DMC $p(y|x)$
if there exists a sequence of $(|\Sc|^n, n)$ joint
source--channel codes such that
\begin{align*}
%\label{eq:di}
\limsup_{n\to \infty} \frac{1}{n} \sum_{i=1}^n \E(d(S_i,\Sh_i)) \le D.
 \end{align*}
%The optimal distortion is the supremum of all achievable distortions $D$.

Shannon's source--channel separation theorem~\cite{Shannon1948, Shannon1959} shows that $D$ is achievable if
\begin{equation} \label{eq:shannon}
 R(D) < C
\end{equation}
and only if $R(D) \le C$, where
$$R(D) = \min_{p(\sh|s):\, \E(d(S,\Sh)) \le D} I(S;\Sh)$$
is the
rate--distortion function for the source $S$ and the distortion measure
$d(s,\sh)$, and
 $$C = \max_{p(x)} I(X;Y)$$
is the capacity of the
channel $p(y|x)$. The proof of achievability is based on separate lossy source coding and channel coding.

In this section, we establish the following alternative characterization of the set of achievable distortions.

\medskip
\begin{theorem}[Shannon~\cite{Shannon1948, Shannon1959}] \label{thm:p2p}
A distortion $D$ is achievable for communication of
the DMS $S$ over the DMC $p(y|x)$
if
\begin{equation} \label{eq:ptp-cond}
 I(S;U) < I(U;Y)
\end{equation}
for some conditional pmf $p(u|s)$, channel encoding function $x(u,s)$, and source decoding function $\sh(u,y)$ such that $\E(d(S,\Sh)) \le D$.
\end{theorem}
\medskip

In the rest of this section, we first describe a joint source--channel coding scheme that is based on the hybrid-coding architecture in Fig.~\ref{fig:arc} and then provide a formal proof of Theorem~\ref{thm:p2p} using the proposed coding scheme.
 %, we first describe the proposed joint source--channel coding scheme based on the hybrid-coding architecture in Fig.~\ref{fig:arc}. %and show that this generalizes both Shannon's separation-based scheme and the uncoded analog transmission strategy.

%Let $d(s,\sh)$  be a non-negative distortion function measuring the cost of representing a symbol $s$ by the symbol $\sh$ from a reconstruction alphabet $\cal \Sh$. The problem is to find sufficient and necessary conditions on the source, the distortion function, and the channel, such that
 %\[
%\frac{1}{n} \sum_{i=1}^n \E(d(S_i,\Sh_i)) \le D.
%\]
%
% so that $S$ and its reconstruction $\hat{S}$ satisfy the fidelity criteria
%\[
%\frac{1}{n} \sum_{i=1}^n \E(d(S_i,\Sh_i)) \le D.
%\]

\subsection{Hybrid Coding Architecture}
\label{sec:hc}
\begin{figure}[h!]
 \centering
\small
\psfrag{s}[B]{$S^n$}
\psfrag{m}[B]{$U^n(M)$}
\psfrag{e1}[c]{Source}
\psfrag{e2}[c]{$x(u,s)$}
\psfrag{d1}[c]{$\sh(u,y)$}
\psfrag{d2}[c]{Channel}
\psfrag{e}[c]{encoder}
\psfrag{p}[c]{$p(y|x)$}
 \psfrag{p2}[b]{$p(y|u)$}
\psfrag{d}[c]{decoder}
\psfrag{x}[B]{$X^n$}
\psfrag{y}[B]{$Y^n$}
\psfrag{mh}[B]{$U^n(\Mh)$}
\psfrag{sh}[B]{$\hat{S}^n$}
\includegraphics[scale=0.365]{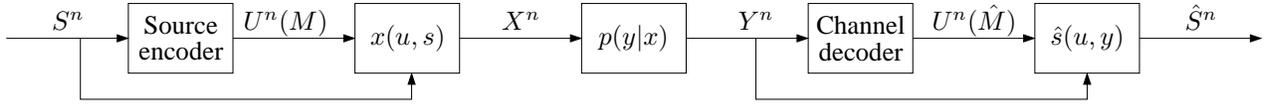}
\caption{The hybrid coding architecture for point-to-point communication.}
 \label{fig:hybrid}
\end{figure}

The proposed joint source--channel coding scheme can be described by the block diagram depicted in Fig.~\ref{fig:hybrid}. A source encoder maps the source sequence $S^n$ into a sequence $U^n(M)$ from a randomly generated codebook $\Cc= \{U^n(m):\, m \in [1:2^{nR}]\}$ of independent and identically distributed codewords. The selected sequence and the source $S^n$ are then mapped {\em symbol-by-symbol} through an encoding function $x(s,u)$ to a sequence $X^n$ that is transmitted over the channel.  Upon receiving the channel output 
sequence $Y^n$, the channel decoder finds an estimate
$U^n(\Mh)$ of $U^n(M)$ and reconstructs $\hat{S}^n$ from $U^n(\Mh)$ and
$Y^n$ again by a {\em symbol-by-symbol} mapping $\sh(u,y)$.

Comparing the general architecture in Fig.~\ref{fig:arc} with the one in Fig.~\ref{fig:hybrid}, we notice that the codeword $U^n(M)$ encodes the (digital) compression index $M \in [1:2^{nR}]$ that has to be reliably transmitted over the channel. At the same time, $U^n(M)$ is the input sequence transmitted over the channel $p(y|u) = \sum_s p(y|x(u,s))p(s|u)$. Hence, $U^n(M)$ plays the roles of both the source codeword that compresses the source sequence within the desired distortion and the channel codeword that encodes the compression index sent across the channel.

The proposed architecture generalizes both Shannon's source--channel separation architecture and the uncoded analog transmission architecture, as shown next.

\begin{figure}[h!]
\centering
\small
\psfrag{s}[B]{$S^n$}
\psfrag{m}[B]{$M$}
\psfrag{e1}[c]{\footnotesize Source}
\psfrag{e2}[c]{\footnotesize Channel}
\psfrag{d1}[c]{\footnotesize Source}
 \psfrag{d2}[c]{\footnotesize Channel}
\psfrag{e3}[c]{}
\psfrag{d3}[c]{}
\psfrag{e}[c]{}
\psfrag{d}[c]{}
\psfrag{e4}[c]{$x(s)$}
\psfrag{d4}[c]{$\sh(y)$}
\psfrag{p}[c]{$p(y|x)$}
\psfrag{x}[B]{$X^n$}
 \psfrag{y}[B]{$Y^n$}
\psfrag{mh}[B]{$\Mh$}
\psfrag{sh}[B]{$\hat{S}^n$}
\includegraphics[scale=0.365]{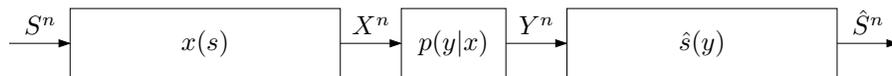}
\caption{Uncoded transmission architecture.}
\label{fig:uncoded}
\end{figure}

\begin{itemize}
\item[a)] {\em Shannon's source--channel separation~\cite{Shannon1948, Shannon1959}}: Under this architecture the source sequence
 is mapped into a compression index $M \in [1:2^{nR}]$, which is then
mapped into a channel codeword $X^n$ to be transmitted over the
channel. Upon receiving $Y^n$, the decoder finds an estimate $\Mh$ of the message
 $M$ and reconstructs $\hat{S}^n(\Mh)$ from $\Mh$.

Suppose that in Theorem~\ref{thm:p2p} we set $U = (X,\Sh)$, where $\Sh\sim p(\sh|s)$ and $X \sim p(x)$ is
independent of $S$ and $\Sh$, $x(u,s)= x$, and $\sh(u,y) = \sh$, so the codeword
 $U^n(M)$ consists of the source codeword $\hat{S}^n(M)$ as well as
the channel codeword $X^n(M)$, the source encoding functions sets the channel input equal to $X^n(M)$, and the source decoding function recovers $\hat{S}^n(\Mh)$ from the estimate $\Mh$ for the transmitted index $M$. In this case the proposed hybrid coding architecture reduces to the source--channel separation
 architecture depicted in Fig.~\ref{fig:separate}. It can be easily checked that~\eqref{eq:ptp-cond} simplifies to $R(D) < C$, that is, Theorem~\ref{thm:p2p} recovers~\eqref{eq:shannon}.

\begin{figure}[h!]
\centering
 \small
\psfrag{s}[B]{$S^n$}
\psfrag{m}[B]{$M$}
\psfrag{e1}[c]{Source}
\psfrag{e2}[c]{Channel}
\psfrag{d1}[c]{Source}
\psfrag{d2}[c]{Channel}
\psfrag{e}[c]{encoder}
\psfrag{p}[c]{$p(y|x)$}
\psfrag{d}[c]{decoder}
 \psfrag{x}[B]{$X^n$}
\psfrag{y}[B]{$Y^n$}
\psfrag{mh}[B]{$\Mh$}
\psfrag{sh}[B]{$\hat{S}^n$}
\includegraphics[scale=0.365]{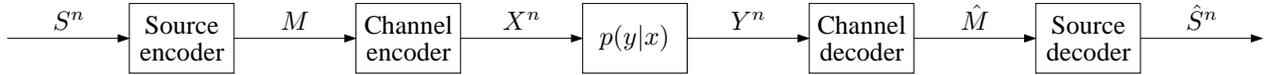}
\caption{Separate source and channel coding system architecture.}
 \label{fig:separate}
\end{figure}

\item[b)] {\em Uncoded transmission}: Under this architecture the source sequence $S^n$ is mapped through a symbol-by-symbol encoding function $x(s)$ into a channel codeword $X^n$ that is transmitted over the
 channel. Upon receiving $Y^n$, the decoder forms an estimate $\hat{S}^n$ of the transmitted source again through a symbol-by-symbol source decoding function $\sh(y)$.
Despite its simplicity, uncoded transmission can be sometimes optimal~\cite{Gastpar2003},
 for example, when communicating a Gaussian source over a Gaussian
channel under the quadratic distortion
measure~\cite{Goblick1965} or communicating a binary source over a
binary symmetric channel under the Hamming distortion measure. In both cases, the desired distortion $D$ can be achieved if $C \ge R(D)$.
 (Note the nonstrict inequality, unlike the strict inequality in Shannon's sufficient
condition~\eqref{eq:shannon}.)

Suppose that in Theorem~\ref{thm:p2p} we set $U = \emptyset$, $x(u,s)= x(s)$, and $\sh(u,y)= \sh(y)$, so the channel input is a symbol-by-symbol function of the source and the source estimate is a  symbol-by-symbol function of the channel output. In this case the proposed hybrid coding architecture reduces to the uncoded transmission architecture depicted in Fig.~\ref{fig:uncoded}, and a distortion $D$ is achievable if there exists $\xh(s)$ and $\sh(y)$ such that
 $\E(d(S,\Sh)) \le D$.
\end{itemize}

The conditions under which a distortion $D$ is achievable can be obtained by studying the conditions for source coding and channel coding separately. Roughly speaking, by
 the lossy source coding theorem, the condition $R > I(U;S)$,
where $R$ is the rate of the codebook $\Cc$,
guarantees a reliable source encoding operation, and by the channel coding
theorem, the condition $R < I(U;Y)$ guarantees a reliable channel
 decoding operation.  Combining the two conditions leads to~\eqref{eq:ptp-cond}.

The precise performance analysis, however,
involves a technical subtlety and requires a careful treatment of the error probability.
 In particular, because $U^n(M)$ is
used as a source codeword, the index $M$ depends on the entire
codebook $\Cc$. But the conventional random coding proof technique for a channel codeword $U^n(M)$ is
developed for situations in which the index $M$ and the (random)
 codebook $\Cc$ are independent of each other. This dependency issue has been well noted by Lapidoth and
Tinguely~\cite[Proof of Proposition D.1]{Lapidoth--Tinguely2010}, who
developed a geometric approach for sending a bivariate Gaussian
 source over a Gaussian multiple access channel. Here we develop a recipe for the general case.

\subsection{Proof of Theorem~\ref{thm:p2p}}

We provide a formal proof of
the sufficient condition~\eqref{eq:ptp-cond} along with a new analysis
 technique that handles the dependency between the transmitted index and the codebook. The standard proof steps are omitted for brevity and can be found in~\cite{El-Gamal--Kim2011}.

\medskip

\emph{Codebook generation:} Let $\e > \e' > 0$. Fix a conditional pmf $p(u|s)$, an encoding function $x(u,s)$, and a source reconstruction function $\sh(u, y)$ such that $\E(d(S,\Sh)) \le D/(1+\e)$. Randomly and independently generate $2^{nR}$ sequences $u^n(m)$,
 $m\in[1:2^{nR}]$, each according to $\prod_{i=1}^n p_U(u_i)$.  The
codebook $\Cc=\{u^n(m):\, m\in[1:2^{nR}]\}$ is revealed to both the
encoder and the decoder.

\emph{Encoding:} We use joint typicality
encoding. Upon observing a sequence $s^n$, the encoder finds an index
 $m$ such that $(u^n(m), s^n)\in \aepvar$. If there is more than one
such index, it chooses one of them at random. If there is no such index,
it chooses an arbitrary index at random from $[1:2^{nR}]$. The encoder
 then transmits $x_i = x(u_i(m),s_i)$ for $i \in [1:n]$.

\emph{Decoding:}
We use joint typicality
decoding. Upon receiving $y^n$, the
decoder finds the unique index $\mh$ such that $(u^n(\mh),y^n)\in
\aep$. If there is none or more than one, it chooses an arbitrary index, say, $\mh = 1$.  The
 decoder then sets the reproduction sequence as $\sh_i=\sh(u_i(\mh),
y_i)$ for $i \in [1:n]$.

\emph{Analysis of the expected distortion:} We bound the distortion averaged over $S^n$, the random choice of
the codebook $\Cc$, and the random index assignment in the encoding procedure.  Let $M$ be the random variable denoting the chosen
 index at the encoder. Define the ``error'' event
\[
\Ec = \bigl\{(S^n, U^n(\Mh), Y^n) \notin \aep\bigr\}
\]
and partition it into
\begin{align*}
\Ec_1 &=\bigl\{(U^n(m), S^n) \notin \aepvar \text{ for all } m\bigr\}, \\
 \Ec_2 &=\bigl\{(S^n, U^n(M), Y^n) \notin \aep\bigr\}, \\
\Ec_3 &=\bigl\{(U^n(m), Y^n)\in \aep \text{ for some } m \ne M \bigr\}.
\end{align*}
Then by the union of events bound,
\begin{align}
\label{eq:p2p_ub}
  \P(\Ec) \le \P(\Ec_1)+\P(\Ec_2\cap \Ec_1^c)+\P(\Ec_3).
\end{align}
We show that all three terms tend to zero as $n \to \infty$ under suitable conditions on the codebook rate $R$. This
implies that the probability of ``error'' tends to zero as $n \to
 \infty$, which, in turn, implies that, by the law of total expectation
and the typical average lemma~\cite[Section~2.4]{El-Gamal--Kim2011},
\begin{align*}
\limsup_{n\to\infty} \E(d(S^n,\hat{S}^n))
&\le \limsup_{n\to\infty} \bigl(\P(\Ec) \E(d(S^n,\hat{S}^n)|\Ec) + \P(\Ec^c) \E(d(S^n,\hat{S}^n)|\Ec^c) \bigr)\\
 &\le (1+\e) \E(d(S,\Sh)),
\end{align*}
and hence the desired distortion is achieved.

By the covering lemma and the conditional typicality lemma
\cite[Sections~2.4 and~3.7]{El-Gamal--Kim2011}, it can be easily
 shown that the first two terms in~\eqref{eq:p2p_ub} tend to zero as $n \to \infty$ if $R >
I(U;S) + \d(\e')$. The third term requires some special attention. By the symmetry of the codebook generation and encoding, we analyze the probability
 conditioned on the event $M = 1$. By the union of events bound, for $n$ sufficiently large,
\begin{align*}
&\P \bigl\{(U^n(m), Y^n)\in \aep \text{ for some } m\neq 1| M=1 \bigr\}\\
&\quad \le \sum_{m=2}^{2^{nR}}\P \bigl\{(U^n(m), Y^n)\in \aep | M=1 \bigr\}\\
 &\quad =\sum_{m=2}^{2^{nR}}\sum_{(u^n, y^n)\in\aep}\P \bigl\{U^n(m)=u^n, Y^n=y^n|M=1 \bigr\}\\
&\quad =\sum_{m=2}^{2^{nR}}\sum_{(u^n, y^n)\in\aep}\P \bigl\{U^n(m)=u^n| Y^n=y^n,M=1 \bigr\}  \P\{Y^n=y^n|M=1\}\\
 %& \qquad \qquad \qquad \cdot \P\{Y^n=y^n|M=1\} \\
%
&\quad = \sum_{m=2}^{2^{nR}}\sum_{(u^n, y^n)\in\aep}
                 \sum_{\ut^n, s^n}  \P\bigl\{U^n(m)=u^n \cond U^n(1) = \ut^n, S^n=s^n ,Y^n=y^n,M=1\bigr\} \\[-1em]
 & \qquad \qquad \qquad \qquad \qquad \quad
\cdot \P\bigl\{U^n(1) = \ut^n, S^n=s^n \cond Y^n=y^n,M=1 \bigr\} \P\{Y^n=y^n|M=1\} \\
&\quad  \stackrel{(a)}{=} \sum_{m=2}^{2^{nR}}\sum_{(u^n, y^n)\in\aep}
               \sum_{\ut^n, s^n} \P\bigl\{U^n(m)=u^n \cond U^n(1) = \ut^n, S^n=s^n ,M=1 \bigr\} \\[-1em]
& \qquad \qquad \qquad \qquad \qquad \quad
\cdot \P\bigl\{U^n(1) = \ut^n, S^n=s^n \cond Y^n=y^n,M=1\bigr\} \P\{Y^n=y^n|M=1\} \\
&\quad \stackrel{(b)}{\le} (1+\e) \sum_{m=2}^{2^{nR}}\sum_{(u^n, y^n)\in\aep}
   \prod_{i=1}^n p_U(u_i) \P\{Y^n=y^n|M=1\} \\
%&\quad \le  (1+\e)\, 2^{nR} \sum_{y^n\in\aep(Y)} \P\{Y^n=y^n|M=1\}  \sum_{u^n\in\aep(U|y^n) } \prod_{i=1}^n p_U(u_i) \\
 &\quad \le  (1+\e)\,  2^{n(R-I(U;Y)+\d(\e))},
\end{align*}
which tends to zero as $n \to \infty$, if $R < I(U;Y) -
\d(\e)$. Here step $(a)$ follows from the fact that given $M=1$, $U^n(m) \to ( U^n(1), S^n )\to Y^n $ form a Markov chain for all $m \neq 1$. To justify step~$(b)$, we make use of the following lemma,
 the proof of which is delegated to Appendix~\ref{app:1}.
%%

%%%%%
\medskip
\begin{lemma}
\label{lem:1}
Let $(U,S)\sim p(u,s)$ and $\e, \e' > 0$. Let $S^n\sim \prod_{i=1}^n p_S(s_i)$
and $U^n(m)$, $m\in[1:2^{nR}]$, be independently generated sequences, each drawn
 according to $\prod_{i=1}^n p_U(u_i)$, independent of $S^n$. Let $\Ic = \{ m \in[1:2^{nR}] :  (U^n(m), S^n ) \in   \aepvar(U,S) \}$ be a set of random indices and let $M \sim \U(\Ic)$, if $|\Ic |$ > 0, and $M \sim \U([1:2^{nR}])$, otherwise.  Then, for every $(u^n,\ut^n,s^n)$,
 \begin{align*}
\P\{U^n(2)=u^n \cond U^n(1)=\ut^n  ,S^n=s^n, M=1\} \le (1+\e) \cdot \prod_{i=1}^n p_{U}(u_i)
\end{align*}
for $n$ sufficiently large.
\end{lemma}
\medskip

%%%%%%
Step $(b)$ above now follows from the inequality in Lemma~\ref{lem:1}, which by symmetry holds for all $\mh \neq 1$. Therefore, if $I(U;S) < I(U;Y) - \d(\e)-\d(\e')$ the probability of ``error'' tends to zero as
 $n\to\infty$ and the average distortion over the random codebook is
bounded as desired. Thus, there exists at least one sequence of codes
achieving the desired distortion. By letting $\e \to 0$, the sufficient
condition~\eqref{eq:ptp-cond} for lossy communication via hybrid coding is established.

\subsection{Discussion}
\label{sec:p2p_disc}

Similar to the source--channel separation architecture, the proposed hybrid coding architecture is modular, whereby the source encoding and channel decoding operations are decoupled and can be analyzed separately. However, unlike the separation architecture, the same codeword is used for both source coding and channel coding, which allows the resulting scheme to perform joint source--channel coding.

The proposed coding scheme can be readily extended to the case of a source transmitted over a DMC with state or over a compound DMC~\cite{FaT}, for which hybrid coding achieves the best known performance, recovering and generalizing several existing results in the literature~\cite{Wyner--Ziv1976,Gelfand--Pinsker1980a,Merhav--Shamai2003,Sutivong-etal-2005,Tuncel2006,Wilson--etal2010,Nayak--Tuncel--Gunduz2010,Gao--Turcel2011b}. The proposed architecture can also be extended to the case of source--channel bandwidth mismatch, whereby $k$ samples of a DMS are transmitted through $n$ uses of a DMC. This can be accomplished by replacing the source and channel symbols in Fig.~\ref{fig:hybrid} by supersymbols of lengths $k$ and $n$, respectively.

\section{Joint Source--Channel Coding over Multiple Access Channels}
\label{sec:networks}

\begin{figure}[h]
\begin{center}
\small
\psfrag{m1}[b]{$S_1^n$}
\psfrag{m2}[b]{$S_2^n$}
\psfrag{x1}[b]{$X_1^n$}
\psfrag{x2}[b]{$X_2^n$}
\psfrag{e1}[c]{Encoder 1}
\psfrag{e2}[c]{Encoder 2}
\psfrag{d}[c]{Decoder}
 \psfrag{p}[c]{$p(y|x_1,x_2)$}
\psfrag{y}[b]{$Y^n$}
\psfrag{mh}[b]{$\hat{S}_1^n,\hat{S}_2^n$}
\includegraphics[scale=0.45]{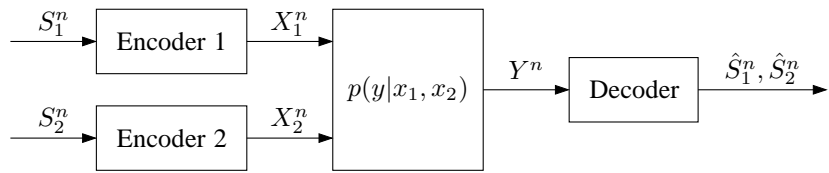}
\end{center}
\caption{Communication of a 2-DMS over a DM-MAC.}
\label{fig:cs-mac}
 \end{figure}

In this section, we illustrate how the hybrid coding system architecture described in Section~\ref{sec:ptp} can be generalized to construct joint source--channel coding schemes for lossy communication over multiuser channels. To illustrate the main ideas, we focus on the specific problem of communicating a pair of correlated discrete
 memoryless sources (2-DMS) $(S_1, S_2)\sim p(s_1,s_2)$ over a discrete memoryless
multiple access channel (DM-MAC) $p(y|x_1, x_2)$, as depicted in Fig.~\ref{fig:cs-mac}. Here each sender $j
= 1,2$ wishes to communicate in $n$ transmissions its source $S_j$ to a common receiver
 so the sources can be reconstructed within desired distortions. %We will consider the block coding setting in which the source sequences $S_1^n = (S_{11},\ldots,S_{1n})$ and $S_2^n = (S_{21}, \ldots, S_{2n})$ are communicated by $n$ transmissions over the channel.

An $(|\Sc_1|^n, |\Sc_2|^n, n)$ joint source--channel code
consists of
\begin{itemize}
\item two encoders, where encoder $j = 1,2$ assigns a sequence
  $x_j^n(s_j^n)\in \Xc_j^n$ to each sequence $s_j^n\in \Sc_j^n$, and
 \item a decoder that assigns an estimate $(\hat s_1^n, \hat s_2^n)\in
  \hat\Sc_1^n\times \hat\Sc_2^n$ to each sequence $y^n\in \Yc^n$.
\end{itemize}
Let $d_1(s_1,\sh_1)$ and $d_2(s_2,\sh_2)$ be two nonnegative distortions measures.
 A distortion pair $(D_1, D_2)$ is said to be achievable
for communication of the 2-DMS $(S_1,S_2)$ over the DM-MAC $p(y|x_1,x_2)$
if there exists a sequence of $(|\Sc_1|^n, |\Sc_2|^n, n)$ joint
source--channel codes such that
 \[
\limsup_{n\to \infty}   \frac{1}{n} \sum_{i=1}^n \E  ( d_j(S_{ji},
\Sh_{ji}) \le D_j, \quad j = 1,2.
\]
The optimal distortion region  is the closure of the set of all
achievable distortion pairs $(D_1,D_2)$.  A computable characterization of the optimal distortion region is not
 known in general. Hybrid coding  yields the following inner bound on the
optimal distortion region.
\medskip
\begin{theorem} \label{thm:mac}
A distortion pair $(D_1, D_2)$ is achievable for communication of
 the 2-DMS $(S_1,S_2)$ over the DM-MAC $p(y|x_1,x_2)$
if
\begin{align*}
I( U_1 ; S_1 | U_2, Q ) &<  I( U_1 ; Y | U_2, Q ), \\
I( U_2 ; S_2 | U_1, Q )  &<  I( U_2 ; Y | U_1, Q ), \\
I( U_1,U_2 ; S_1, S_2 | Q )    &<  I( U_1 , U_2 ; Y |Q)
 \end{align*}
for some pmf $p(q) p(u_1|s_1,q)p(u_2|s_2,q)$ and functions $x_1(q,u_1,s_1)$, $x_2(q,u_2,s_2)$, $\sh_1(q,u_1,u_2,y)$, and $\sh_2(q,u_1,u_2,y)$ such that
$\E(d_j(S_j,\Sh_j)) \le D_j$, $j=1,2$.
\end{theorem}
 \medskip
The proof of the theorem is given in
Appendix~\ref{sec:mac_achie}.

%--------------------------------------------------------------------
%\subsubsection{Special Cases}
%\label{sec:special_MAC}
 %--------------------------------------------------------------------

Application of Theorem~\ref{thm:mac} yields the following
results as special cases:
\begin{itemize}
%\subsubsection{Lossless Communication} \label{subsec:lossless}
 \item[a)] {\em Lossless communication}: When specialized to the case in which $d_1$ and $d_2$ are Hamming
distortion measures
%\begin{equation*}
%d_j(s_j,\sh_j) =
%\begin{cases}
%1, & \text{if } \sh_j = s_j,\\
 %0, & \text{if } \sh_j \neq s_j,
%\end{cases}
%\qquad  j=1,2,
%%\label{eq:hamming}
%\end{equation*}
and $D_1 = D_2 = 0$,
Theorem~\ref{thm:mac} recovers the following sufficient condition for lossless
 communication of a 2-DMS over a DM-MAC.

\medskip
\begin{corollary}[Cover, El Gamal, and Salehi~\cite{Cover--El-Gamal--Salehi1980}]
\label{cor:ces}
A 2-DMS $(S_1, S_2)$ can be communicated losslessly over the DM-MAC
 $p(y|x_1, x_2)$ if
\begin{align*}
H(S_1| S_2) &< I(X_1; Y|X_2, S_2, Q), \\
H(S_2| S_1) &< I(X_2; Y|X_1, S_1, Q), \\
H(S_1, S_2) &< I(X_1, X_2 ; Y|Q)
\end{align*}
for some pmf $p(q)p(x_1|s_1,q)p(x_2|s_2,q)$.
 \end{corollary}
\medskip
\begin{IEEEproof}
It suffices to choose in Theorem~\ref{thm:mac} $U_j = (X_j,S_j)$, $x_j(q,u_j,s_j) = x_j$, and $\sh_j(q,u_1,u_2,y) = s_j$, $j =1,2$, under a pmf of the form $p(q)p(x_1|s_1,q)p(x_2|s_2,q)$.
 \end{IEEEproof}
\medskip

%--------------------------------------------------------------------
%\subsubsection{Distributed Lossy Source Coding} \label{subsec:dlsc}
\item[b)] {\em Distributed lossy source coding}: When specialized to the case of a noiseless DM-MAC $Y = (X_1,X_2)$ with $\log|\Xc_1|
 = R_1$ and $\log|\Xc_2| = R_2$ and $(X_1,X_2)$ independent of the sources, Theorem~\ref{thm:mac} recovers the Berger--Tung inner
bound on the rate--distortion region for distributed lossy source
coding.

%Wagner~{\it et al.}~\cite{Wagner--Kelly--Altug2011},
 \medskip
\begin{corollary}[Berger~\cite{Berger1978} and
Tung~\cite{Tung1978}]\label{cor:bt}
A distortion pair $(D_1, D_2)$ with rate pair $(R_1,R_2)$ is achievable for distributed lossy source coding of a 2-DMS $(S_1,S_2)$ if
 \begin{align*}
R_1 &> I(S_1; U_1|U_2,Q),\\
R_2 &> I(S_2; U_2|U_1,Q),\\
R_1+R_2 &> I(S_1, S_2; U_1, U_2|Q)
\end{align*}
for some pmf $p(q)p(u_1|s_1,q)p(u_2|s_2,q)$
and functions $\sh_1(q,u_1,u_2)$ and $\sh_2(q,u_1,u_2)$
 such that $\E(d_j(S_j, \Sh_j)) \le D_j$, $j =1,2$.
\end{corollary}
\medskip
\begin{IEEEproof}
It suffices to choose in Theorem~\ref{thm:mac} $U_j = (X_j,\Ut_j)$, $x_j(q,u_j,s_j) = x_j$, and $\sh_j(q,u_1,u_2,y) = \sh_j(q,\ut_1,\ut_2)$, $j =1,2$,  under a pmf of the form $p(q)p(\ut_1|s_1,q)$ $p(\ut_2|s_2,q)$, and to relabel the tilded random variables.
 \end{IEEEproof}
\medskip

%\subsubsection{Bivariate Gaussian source over a Gaussian MAC} \label{subsec:MAC_Gaussian}

\item[c)] {\em Bivariate Gaussian source over a Gaussian MAC}: Suppose that the source is a bivariate Gaussian pair with equal variance $\sigma^2$
  and that each source component has to be reconstructed by the decoder under quadratic distortion measures
$d_j(s_j,\sh_j) = (s_j-\sh_j)^2$, $j=1,2$. In addition, assume that the channel is the Gaussian MAC  $Y = X_1 + X_2 + Z$,  where $Z$ is AWGN and the channel inputs $X_1$ and $X_2$ are subject to average power constraints.  Theorem~\ref{thm:mac} can be adapted to this case via the
 standard discretization method~\cite[Sections~3.4 and~3.8]{El-Gamal--Kim2011}.
  Suppose that in Theorem~\ref{thm:mac} we choose $(U_1,U_2)$ as jointly Gaussian random variables conditionally independent given $(S_1,S_2)$, the encoding function $x_j(u_j,s_j)$, $j=1,2$, as a linear function of $u_j$ and $s_j$, and the decoding function $\sh_j(u_1,u_2,y)$ as the minimum mean-square error (MMSE) estimate of $S_j$ given $U_1$, $U_2$, and $Y$. Then, Theorem~\ref{thm:mac} recovers to the sufficient
 condition for lossy communication derived in Lapidoth and Tinguely~\cite[Theorem IV.6]{Lapidoth--Tinguely2010} via a hybrid analog/digital scheme that combines uncoded transmission and vector quantization.
\end{itemize}

%%%%
%%%%
\subsection{Hybrid Coding Architecture}
 \label{sec:mac_hybrid}

\begin{figure}[h]
 \begin{center}
\small
\psfrag{s}[B]{$S_1^n$}
\psfrag{s2}[B]{$S_2^n$}
\psfrag{m}[B]{$U_1^n(M_1)$}
\psfrag{m2}[B]{$U_2^n(M_2)$}
\psfrag{e1}[c]{Source}
\psfrag{e12}[c]{Source}
\psfrag{e2}[c]{$x_1(u_1,s_1)$}
 \psfrag{e22}[c]{$x_2(u_2,s_2)$}
\psfrag{d1}[c]{$\sh_1(u_1,u_2,y)$}
\psfrag{d12}[c]{$\sh_2(u_1,u_2,y)$}
\psfrag{d2}[c]{Channel}
\psfrag{e}[c]{encoder 1}
\psfrag{e02}[c]{encoder 2}
\psfrag{p}[c]{$p(y|x_1,x_2)$}
 \psfrag{p2}[b]{$p(y|u_1,u_2)$}
\psfrag{d}[c]{decoder}
\psfrag{x}[B]{$X_1^n$}
\psfrag{x2}[B]{$X_2^n$}
\psfrag{y}[B]{$Y^n$}
\psfrag{mh}[B]{$U_1^n(\Mh_1)$}
\psfrag{mh2}[B]{$U_2^n(\Mh_2)$}
\psfrag{sh}[B]{$\hat{S}_1^n$}
 \psfrag{sh2}[B]{$\hat{S}_2^n$}
\includegraphics[scale=0.49]{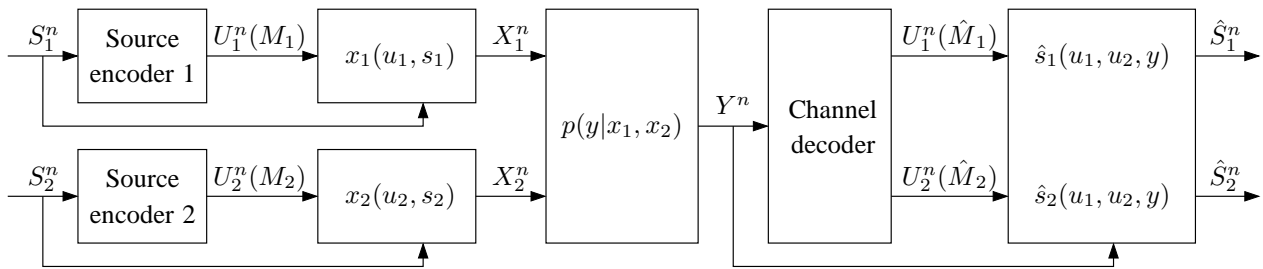}
\end{center}
\caption{Joint source--channel coding system architecture for
  communicating a 2-DMS over a DM-MAC.}
\label{fig:newmac}
 \end{figure}

The joint source--channel coding scheme used in the proof of achievability of  Theorem~\ref{thm:mac} is based on the hybrid coding system architecture depicted in Fig.~\ref{fig:newmac}.
Here the source sequence $S^n_j$ is mapped by source encoder $j=1,2$ into a sequence $U_j^n(M_j)$ from a randomly generated codebook $\Cc_j = \{ U_j^n(m_j): m_j \in [1:2^{nR_j}] \}$ of independently distributed codewords. The selected sequence and the source $S^n_j$ are then mapped
 symbol-by-symbol through an encoding function $x_j(s_j,u_j)$ to a sequence $X_j^n$, which is transmitted over the MAC. Upon receiving  $Y^n$, the decoder finds the estimates $U_1^n(\Mh_1)$ and $U_2^n(\Mh_2)$ of $U_1^n(M_1)$ and $U_2^n(M_2)$, respectively, and reconstructs $\hat{S}_1^n$ and $\hat{S}_2^n$ from $U_1^n(\Mh_1)$, $U_2^n(\Mh_2)$, and $Y^n$ by symbol-by-symbol mappings $\sh_j(u_1,u_2,y)$, $j=1,2$. Comparing Fig.~\ref{fig:newmac} with Fig.~\ref{fig:hybrid}, we notice that the two separate encoders in Fig.~\ref{fig:newmac} operate exactly as the hybrid encoder in Fig.~\ref{fig:hybrid}, while the channel decoder at the output of the DM-MAC differs from the one used in the point-to-point case because it {\em jointly} decodes $U_1^n(\Mh_1)$ and $U_2^n(\Mh_2)$ and forms the source estimates from the pair $(U_1^n(\Mh_1),U_2^n(\Mh_2))$ as well as $Y^n$.

As in the case of point-to-point communication, the conditions under which a distortion pair $(D_1,D_2)$ is achievable can be obtained by studying the conditions for source coding and channel coding separately. By the covering lemma~\cite[Section~3.7]{El-Gamal--Kim2011}, the source encoding
 operation is successful if
\begin{align*} %
R_1 &> I(U_1;S_1),\\
R_2 &> I(U_2;S_2),
\end{align*}
while by the packing lemma~\cite[Section~3.2]{El-Gamal--Kim2011}, suitably modified to account for the dependence between the indices and the codebook that we have mentioned in Section~\ref{sec:ptp}, the channel decoding operation is successful if
 \begin{align*}
R_1&<I(U_1;Y,U_2),\\
R_2&<I(U_2;Y,U_1),\\
R_1+R_2 &< I(U_1,U_2;Y) + I(U_1;U_2).
\end{align*}
Then, the sufficient condition in Theorem~\ref{thm:mac} (with $Q = \emptyset$) is established by combining the above inequalities and eliminating the intermediate rate pair $(R_1,R_2)$. The sufficient condition with a general $Q$ can be proved by introducing a time sharing random variable~$Q$
 and using coded time sharing~\cite[Section~4.5.3]{El-Gamal--Kim2011}.

\subsection{Remarks}
\label{sec:mac_discussion}

The proposed joint source--channel coding scheme is conceptually similar to separate source and channel coding and, loosely speaking, is obtained by concatenating the source coding scheme in~\cite{Berger1978,Tung1978} for distributed lossy source coding with a channel code for multiple access communication, except that the same codeword is used by both the source encoder and the channel encoder.
%each encoder for source coding as well as for channel coding. 
%Theorem~\ref{thm:mac} generalizes the coding scheme by Cover, El Gamal, and Salehi for lossless communication over a DM-MAC~\cite{Cover--El-Gamal--Salehi1980} to the case of lossy communication. 
Similarly to the coding scheme by Cover, El Gamal, and Salehi in~\cite{Cover--El-Gamal--Salehi1980} for lossless communication over a DM-MAC, the hybrid coding in Theorem~\ref{thm:mac} enables coherent communication over the MAC by preserving the correlation between the sources at the channel inputs chosen by the two senders.

%WITHIN A CONCEPTUALLY FAMILIAR FRAMEWORK, HYBRID CODING ENABLES COHERENT COMMUNICATION OVER THE MULTIPLE ACCESS CHANNEL BY CONVEYING THE SOURCE CORRELATION TO THE CHANNEL INPUTS. PRESERVING THE CORRELATION BETWEEN THE SOURCES TO THE CHANNEL INPUTS IS REMINISCENT OF COVER, EL GAMAL, AND SALEHI'S SCHEME FOR THE LOSSLESS CASE~\cite{Cover--El-Gamal--Salehi1980} AND HYBRID CODING CAN BE CONSIDERED AS ITS GENERALIZATION TO THE LOSSY CASE.

The achievable distortion region in Theorem~\ref{thm:mac} can be increased when the 2-DMS $(S_1,S_2)$ has a nontrivial common part in the sense of G\'acs--K\"orner~\cite{Gacs--Korner1973} and Witsenhausen~\cite{Witsenhausen1975}. In this case, the encoders can jointly compress the common part and use it to establish coherent communication over the MAC. This extension will be considered elsewhere~\cite{FaT}, where a hybrid coding scheme is proposed by combining the distributed lossy source coding scheme in~\cite{Wagner--Kelly--Altug2011} for sources with a nonempty common part and the channel coding scheme in~\cite{Slepian--Wolf1973b} for multiple access communication with a common message shared by the two encoders. The result in Theorem~\ref{thm:mac} can also be generalized to the setting in which the source consists of a random triple $(S,S_1,S_2)$, the distortion measures are $d_j: \Sc \times \Sc_1 \times \Sc_2 \to \hat{\Sc}_j$, $j=1,2$, but encoder~$j$ can only observe the source component $S_j$, $j=1,2$. This setting includes as special cases the CEO problem~\cite{Berger--Zhang--Viswanathan1996,Viswanathan--Berger1997,Prabhakaran--Tse--Ramchandran2004,Oohama2005} and the Gaussian sensor network~\cite{Gastpar2008}.

The modular approach presented here for lossy communications over multiple access channels can be adapted to construct joint source--channel coding schemes for other channel models. In~\cite{FaT}, extensions to several canonical channel models studied in the literature will be
 presented---the broadcast channel, the interference channel, as well as channels with state or noiseless output feedback. In all these examples, we establish sufficient conditions for lossy communication based on hybrid coding. The basic design principle consists in combining a source coding scheme with a suitably ``matched'' channel coding scheme by the means of the hybrid coding architecture described in Section~\ref{sec:ptp}. For instance, in the case of lossy communication over broadcast channels, a hybrid coding scheme can be constructed by concatenating the Gray--Wyner lossy source coding scheme \cite{Gray--Wyner1974} with the Marton coding scheme \cite{Marton1979} for the general broadcast channel with a common message.

%%%%%%%%%%%%%%%%%%%%%%%%%%%%%%%%%
%%%%%%%%%%%%%%%%%%%%%%%%%%%%%%%%%
%%
%%%%%%%%    RELAY NETWORKS     %%%%%%%%%%%%
%%
%%%%%%%%%%%%%%%%%%%%%%%%%%%%%%%%%%
%%%%%%%%%%%%%%%%%%%%%%%%%%%%%%%%%%

\section{Relay Networks}
\label{sec:relay}

In this section we explore applications of hybrid coding in the context of relay networks, wherein a source node wishes to send a message to a destination node with the help of intermediate relay nodes. Over the past decades, three dominant paradigms have been proposed for  relay communication: decode--forward, compress--forward, and amplify--forward.
 \begin{itemize}
\item In decode--forward, each relay  recovers the transmitted message by the source either fully or partially and forwards it to the receiver (digital-to-digital
interface) while
coherently cooperating with the source node.
 Decode--forward was originally proposed in~\cite{Cover--El-Gamal1979}  for the relay channel and has been generalized to multiple relay networks, for example, in~\cite{Aref1980, Kramer--Gastpar--Gupta2005} and further improved by
 combining it with structured coding \cite{Nazer--Gastpar2007, Nam--Chung--Lee2009}.

\item In amplify--forward, each relay sends a scaled version of its received sequence and forwards it to the receiver (analog-to-analog interface). Amplify--forward was  proposed in~\cite{schein--gallager2000} for the Gaussian two--relay diamond network and subsequently studied for the Gaussian relay channel in~\cite{Laneman-etal2004}. Generalizations of amply--forward to general nonlinear analog mappings for relay communication have been proposed in~\cite{Khormuji--Skoglund2010}.

\item In compress--forward, each relay vector-quantizes its received sequence and forwards it to the receiver (analog-to-digital interface).
Compress--forward was proposed in~\cite{Cover--El-Gamal1979}  for the relay channel and has been generalized to arbitrary noisy networks in~\cite{Lim--Kim--El-Gamal--Chung2011} as noisy network coding. % which also extends network coding~\cite{Ahlswede--Cai--Li--Yeung2000}.
 \end{itemize}

In this section we propose a new coding scheme for relay networks that uses hybrid analog/digital coding at the relay nodes. The proposed scheme naturally extends both amplify--forward and compress--forward since each relay node uses the hybrid coding architecture introduced in Section~\ref{sec:ptp} to transmit a symbol-by-symbol function of the received sequence and its quantized version (analog-to-analog/digital interface). More important than this conceptual unification is the performance improvement of hybrid coding. We demonstrate through two specific examples, the two--way relay channel (Section~\ref{sec:twrc}) and the two--relay diamond network (Section~\ref{sec:diamond}), that hybrid coding can strictly outperform the existing coding schemes, not only amplify--forward and compress--forward, but also decode--forward.

\subsection{Two--Way Relay Channel}\label{sec:twrc}

\begin{figure}[h]
\begin{center}
\small
\psfrag{m1}[b]{$M_1$}
\psfrag{m2}[b]{$M_2$}
\psfrag{mh1}[b]{$\Mh_1$}
\psfrag{mh2}[b]{$\Mh_2$}
\psfrag{x1}[b]{$X_1^n$}
\psfrag{x2}[b]{$X_2^n$}
\psfrag{x3}[b]{$X_3^n$}
 \psfrag{p1}[c]{$p(y_3|x_1,x_2)$}
\psfrag{p2}[c]{$p(y_1,y_2|x_3)$}
\psfrag{y1}[b]{$Y_1^n$}
\psfrag{y2}[b]{$Y_2^n$}
\psfrag{y3}[b]{$Y_3^n$}
\psfrag{1}[c]{1}
\psfrag{2}[c]{2}
\psfrag{3}[c]{3}
\includegraphics[scale=0.45]{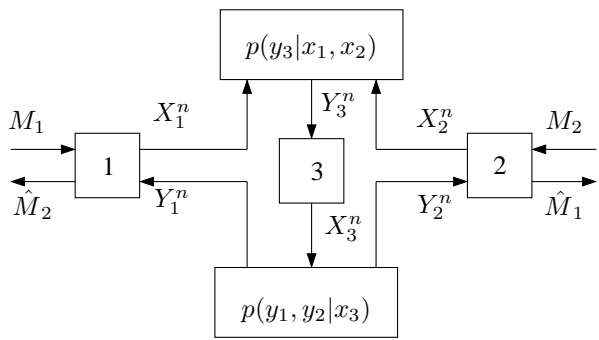}
 \end{center}
\caption{The two--way relay channel.}
\label{fig:twrc}
\end{figure}

Consider the relay network depicted in Fig.~\ref{fig:twrc}, where
two source/destination nodes communicate with each other with the help of one relay. Node~1 wishes to send the message $M_1 \in [1:2^{nR_1}]$ to node~2 and node~2 wishes to send the message $M_2 \in
 [1:2^{nR_2}]$ to node~1 with the help of the relay node~3. Nodes 1 and 2 are connected to the relay through the MAC $p(y_3|x_1,x_2)$, while the relay is connected to nodes 1 and 2 via the broadcast channel $p(y_1,y_2|x_3)$.

This network is modeled by a 3-node discrete memoryless two--way relay channel (DM-TWRC) $p(y_1,y_2|x_3) p(y_3|x_1,x_2)$. A $(2^{nR_1},2^{nR_2},n)$ code for the DM-TWRC consists of
 \begin{itemize}
\item two message sets $[1:2^{nR_1}]\times [1:2^{nR_2}]$,
\item two encoders, where at time $i \in [1:n]$ encoder $j = 1,2$ assigns a symbol $x_{j,i}(m_j, y_j^{i-1}) \in \Xc_j$ to each message $m_j \in [1:2^{nR_j}]$ and past received output sequence $y_j^{i-1} \in \Yc_j^{i-1}$,
 \item a relay encoder that assigns a symbol $x_{3,i}( y_3^{i-1})$ to each past received output sequence $y_3^{i-1} \in \Yc_3^{i-1}$, and
\item two decoders, where decoder 1 assigns an estimate $\mh_2$ or an error message to each received sequence $y_1^n \in \Yc_1^n$ and decoder 2 assigns an estimate $\mh_1$ or an error message to each received sequence $y_2^n \in \Yc_2^n$.
 \end{itemize}
We assume that the message pair $(M_1,M_2)$ is uniformly distributed over
$[1:2^{nR_1}]\times [1:2^{nR_2}]$. The average probability of error is defined as $P_e^{(n)} = \P\{ (\Mh_1,\Mh_2) \ne (M_1,M_2) \}$. A rate pair $(R_1,R_2)$ is said to be achievable for the DM-TWRC if there exists a sequence of $(2^{nR_1},2^{nR_2},n)$ codes such that $\lim_{n \to \infty} P_e^{(n)} = 0$. The capacity region of the DM-TWRC is the closure of the set of achievable rate pairs $(R_1,R_2)$ and the sum-capacity is the supremum of the achievable sum rates $R_1+R_2$.

The capacity region of the DM-TWRC $p(y_1,y_2|x_3)p(y_3|x_1,x_2)$ is not known in general. Rankov and Wittneben~\cite{Rankov--Wittneben2006} characterized inner bounds on the capacity region based on decode--forward,  compress--forward, and amplify--forward.
 Another inner bound based on noisy network coding is given in~\cite{Lim--Kim--El-Gamal--Chung2011}.  In the special case of a Gaussian TWRC, Nam, Chung, and Lee~\cite{Nam--Chung--Lee2009} proposed a coding scheme based on nested lattice codes and structured binning that achieves within $1/2$ bit per dimension from the capacity region for all underlying channel parameters.

%Most of the existing inner bounds~\cite{Rankov--Wittneben2006} are based on classical coding techniques originally developed for the relay channel, such as decode--forward, compress--forward, amplify--forward, compute--forward, noisy network coding, or a combination of these.
 %Such techniques yield the capacity region~\cite{kang--ulukus}, or an approximation of it~\cite{urs--diggavi}, for special classes of channels.
%We notice, however, that if the channel input sequences $(X^n_1,X^n_2)$ are i.i.d.\@ $\sim p(x_1)p(x_2)$, then the channel output sequence $Y_3^n$ observed the relay is a DM ``source'' $\sim \sum_{x_1,x_2}p(y_3|x_1,x_2)p(x_1)p(x_2)$, which can be naturally
 %communicated from the relay to the destinations using joint source--channel coding techniques.

Hybrid coding yields the following inner bound on the capacity region, the proof of which is given in  Appendix~\ref{sec:proof-twrc}.

\medskip
\begin{theorem}
\label{thm:dm-twrc}
A rate pair $(R_1, R_2)$ is achievable for the DM-TWRC $p(y_1,y_2|x_3) p(y_3|x_1,x_2)$ if
\begin{equation} \label{eq:dm-twrc}
%\begin{split}
%R_1 &< I(X_1;Y_2, U_3|X_2),\\
 %R_1 &< I(X_1,U_3; X_2,Y_2) + I(X_1;U_3) - I(Y_3;U_3),\\
%R_2 &< I(X_2;Y_1, U_3|X_1),\\
%R_2 &< I(X_2,U_3; X_1,Y_1) + I(X_2;U_3) - I(Y_3;U_3)
%\end{split}
\begin{split}
R_1 &< \min\bigl( I(X_1;Y_2, U_3|X_2),I(X_1,U_3; X_2,Y_2) - I(Y_3;U_3|X_1) \bigr),\\
 %R_1 &< I(X_1,U_3; X_2,Y_2) - I(Y_3;U_3|X_1),\\
R_2 &< \min \bigl( I(X_2;Y_1, U_3|X_1), I(X_2,U_3; X_1,Y_1) - I(Y_3;U_3|X_1) \bigr) ,
%\\ R_2 &< I(X_2,U_3; X_1,Y_1) - I(Y_3;U_3|X_1)
\end{split}
 \end{equation}
for some pmf $p(x_1)p(x_2)p(u_3|y_3)$ and function
$x_3(u_3,y_3)$.
\end{theorem}

\medskip

%%%
\begin{remark}
Theorem~\ref{thm:dm-twrc} includes both the noisy network coding inner bound, which is recovered by letting $U_3 = (\Yh_3, X_3)$ under a pmf $p(\yh_3,x_3|y_3) =
 p(\yh_3|y_3)p(x_3)$, and the amplify--forward inner bound, which is obtained by setting $U_3 = \emptyset$, and the inclusion can be strict in general.
\end{remark}

\subsubsection{Gaussian Two-Way Relay Channel}
 As an application of Theorem~\ref{thm:dm-twrc},
consider the special case of the Gaussian TWRC, where the channel outputs corresponding to the inputs $X_1, X_2,$ and  $X_3$ are
\begin{align*}
Y_1 &= g_{13} X_3 + Z_1,\\
 Y_2 &= g_{23} X_3 + Z_2,\\
Y_3 &= g_{31} X_1 + g_{32} X_2 + Z_3,
\end{align*}
and the noise components $Z_k$, $k=1,2,3$, are i.i.d.\@ $\N(0,1)$. The channel gains $g_{kj}$ from node $j$ to node $k$ are assumed to be real, constant over time, and known throughout the network. We assume expected power constraint $P$ at each sender.  Denote the received SNR $S_{jk} = g_{jk}^2 P$. %The capacity region of the Gaussian TWRC is not known in general.

Theorem~\ref{thm:dm-twrc} yields the following inner bound on the capacity region.
\medskip
\begin{corollary}
\label{cor:AWGN-TWRC}
A rate pair $(R_1, R_2)$ is achievable for the Gaussian TWRC if
\begin{align*}
 R_1 &< \half \log
       \frac{\left(\frac{\a S_{23}(S_{31}+1)}{S_{31}+S_{32}+1}
                   +\b S_{23}+1 \right)(S_{31}+1+\sigma^2)
             - S_{23}\left(\sqrt{\frac{\a(S_{31}+1)}{S_{31}+S_{32}+1}}
                            + \sqrt{\b\sigma^2}\right)^2}
       {\left(\frac{\a S_{23}}{S_{31}+S_{32}+1}
                   +\b S_{23}+1 \right)(1+\sigma^2)
             - S_{23}\left(\sqrt{\frac{\a}{S_{31}+S_{32}+1}}
                            + \sqrt{\b\sigma^2}\right)^2},\\
R_1 &< \half \log
       \frac{\left(\frac{\a S_{23}(S_{31}+1)}
                   {S_{31}+S_{32}+1} + (1-\a) S_{23} + 1 \right)
             (1+\sigma^2)}
        {\left(\frac{\a S_{23}}{S_{31}+S_{32}+1} + \b S_{23} + 1\right)
        (1+\sigma^2) - S_{23} \left(\sqrt{\frac{\a}{S_{31}+S_{32}+1}}
                           + \sqrt{\b\sigma^2}\right)^2} - \C(1/\sigma^2),\\
 R_2 &< \half \log
       \frac{\left(\frac{\a S_{13}(S_{32}+1)}{S_{31}+S_{32}+1}
                   +\b S_{13}+1 \right)(S_{32}+1+\sigma^2)
             - S_{13}\left(\sqrt{\frac{\a(S_{32}+1)}{S_{31}+S_{32}+1}}
                            + \sqrt{\b\sigma^2}\right)^2}
       {\left(\frac{\a S_{13}}{S_{31}+S_{32}+1}
                   +\b S_{13}+1 \right)(1+\sigma^2)
             - S_{13}\left(\sqrt{\frac{\a}{S_{31}+S_{32}+1}}
                            + \sqrt{\b\sigma^2}\right)^2},\\
R_2 &< \half \log
       \frac{\left(\frac{\a S_{13}(S_{32}+1)}
                   {S_{31}+S_{32}+1} + (1-\a) S_{13} + 1 \right)
             (1+\sigma^2)}
        {\left(\frac{\a S_{13}}{S_{31}+S_{32}+1} + \b S_{13} + 1\right)
        (1+\sigma^2) - S_{13} \left(\sqrt{\frac{\a}{S_{31}+S_{32}+1}}
                           + \sqrt{\b\sigma^2}\right)^2} - \C(1/\sigma^2)
 \end{align*}
for some $\a,\b \in [0,1]$ such that $\a + \b \le 1$ and $\sigma^2 > 0$.
\end{corollary}
\medskip
\begin{IEEEproof}
It suffices to set in Theorem~\ref{thm:dm-twrc} $X_1$ and $X_2$ as i.i.d. $\sim \N(0,P)$, $U_3 = (V_3,\Yh_3)$, where $\Yh_3 = Y_3+\Zh_3$,
 $\Zh_3$ and $V_3$ are i.i.d. zero-mean Gaussian independent of $(X_1,X_2,Y_3)$ with variance $\sigma^2$ and $1$, respectively, and
\begin{equation}
\label{eq:twrc_x}
x_3(u_3,y_3) = \sqrt{\frac{\a P}{S_{31} + S_{32} + 1}}\, y_3
     + \sqrt{\frac{\b P}{\sigma^2}}\, (y_3 - \yh_3)
    + \sqrt{(1-\a-\b) P}\, v_3,
\end{equation}
for some $\a,\b \in [0,1]$ such that $\a + \b \le 1$ and $\E(X^2_3) =P$.
\end{IEEEproof}

\medskip
Note from~\eqref{eq:twrc_x} that the channel input sequence produced by the relay node is a linear combination of the (analog) sequence $Y_3$, the 
(digital) quantized sequence $\Yh_3=Y_3+\Zh_3$, whose resolution is determined by~$\sigma_j^2$, and the (digital) sequence $V_3$.
%the (digital/analog) error $\Zh_3= Y_3 - \Yh_3$ of the quantized sequence $\Yh_3=Y_3+\Zh_3$, whose resolution is determined by~$\sigma_j^2$, and the (digital) sequence $V_3$. %THE (DIGITAL) INDEX OF THE COMPRESSION SEQUENCE $\Yh_3$ IS CARRIED BY BOTH ITSELF AND THE SEQUENCE $V_3$ WHICH IS CONSISTENT WITH THE FACT THAT WE SET $U_3=(V_3, \Yh_3)$. 
Hence by varying $\a$ and $\b$, we can vary the amount of power allocated to the digital and analog parts in order to optimize the achievable rate region. In particular, by letting $\a=\b=0$, then the hybrid coding inner bound in Corollary~\ref{cor:AWGN-TWRC} reduces to the noisy network coding inner bound~\cite{Lim--Kim--El-Gamal--Chung2011}  that consists of all rate pairs $(R_1,R_2)$ such that
 \begin{equation}
\begin{split}
R_1 &< \min\left( \C\left( \tfrac{S_{31}  }{ 1 + \sigma^2 } \right), \C( S_{23} ) - \C(1/\sigma^2) \right),\\
R_2 &< \min\left( \C\left( \tfrac{S_{32}  }{ 1 + \sigma^2 } \right), \C( S_{13} ) - \C(1/\sigma^2) \right)
 \end{split}
\label{eq:TWRCnnc}
\end{equation}
for some $\sigma^2 > 0$. If instead we let $\a = 1$, $\b =
0$, and $\sigma^2 \to \infty$, then the hybrid coding inner bound
reduces to the
amplify--forward inner bound~\cite{Rankov--Wittneben2006} that consists of all rate pairs $(R_1,R_2)$ such that
 \begin{equation}
\begin{split}
R_1 &<  \C\left( \frac{ S_{23}  S_{31}  }{ 1 + S_{23} + S_{31} + S_{32} } \right),\\
R_2 &<  \C\left( \frac{ S_{13} S_{32}  }{ 1 + S_{13} + S_{31} + S_{32} } \right).
 \end{split}
\label{eq:TWRCaf}
\end{equation}
Similarly, by letting $\a = 0$ and  $\b = 1$, then the hybrid coding inner bound in Corollary~\ref{cor:AWGN-TWRC}  reduces to the set of rate pairs $(R_1,R_2)$ such that
 \begin{equation}
\begin{split}
R_1 &< \min\left( \C\left( \frac{S_{31}(1+S_{23}) }{ 1 + \sigma^2 + S_{23} } \right), \C\left( \tfrac{S_{23} \sigma^2 }{ 1 + \sigma^2 + S_{23} } \right) - \C(1/\sigma^2) \right),\\
 R_2 &< \min \left( \C\left( \frac{S_{32}(1+S_{13}) }{ 1 + \sigma^2 + S_{13} } \right),  \C\left( \frac{S_{13} \sigma^2 }{ 1 + \sigma^2 + S_{13} } \right) - \C(1/\sigma^2) \right)
\end{split}
\label{eq:TWRCnew}
 \end{equation}
for some $\sigma^2 > 0$. Finally, by setting $\alpha+\beta=1$, Corollary~\ref{cor:AWGN-TWRC} includes as a special case a hybrid coding scheme recently proposed in~\cite{Khormuji--Skoglund2011}.

\begin{figure}[h]
\centering
\footnotesize
\psfrag{r}[l]{$r$}
\psfrag{Rcut}[bl]{$R_\mathrm{CS}$}
\psfrag{Raf}[bl]{$R_\mathrm{AF}$}
\psfrag{Rdf}[bl]{$R_\mathrm{DF}$}
 \psfrag{Rnnc}[bl]{$R_\mathrm{NNC}$}
\psfrag{Rhc}[bl]{$R_\mathrm{HC}$}
\hspace*{-1.5em}\includegraphics[scale=0.6]{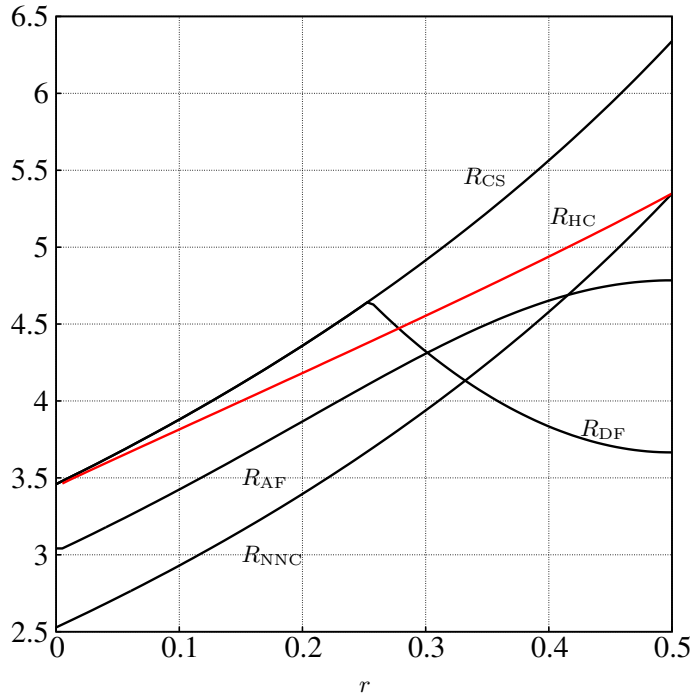}
\caption{Comparison of the cutset bound $R_\mathrm{CS}$,
  decode--forward lower bound $\R_\mathrm{DF}$, amplify--forward lower
   bound $R_\mathrm{AF}$, noisy network coding lower bound
  $R_\mathrm{NNC}$, and hybrid coding lower bound $R_\mathrm{HC}$
   on the sum-capacity for the Gaussian TWRC as a function of the distance $r$ between nodes 1 and 3.}
 \label{fig:gnets-2waycomp}
\end{figure}

Fig.~\ref{fig:gnets-2waycomp} compares the cutset bound~\cite{El-Gamal--Kim2011} on
 the sum-capacity with the inner bound achieved by decode--forward~\cite{Rankov--Wittneben2006},  noisy network coding~\eqref{eq:TWRCnnc}, amplify--forward~\eqref{eq:TWRCaf}, and hybrid coding~\eqref{eq:TWRCnew}. The plots in the figure assume that nodes~1 and~2 are unit distance apart and
node~3 is at distance $r \in [0,1]$ from node~1 along the line between
nodes~1 and~2; the channel gains are of the form
 $g_{jk} = r_{jk}^{-3/2}$, where $r_{jk}$ is the distance between
nodes $j$ and $k$, hence $g_{13} =g_{31} = r^{-3/2}$, $g_{23} =
g_{32}= (1-r)^{-3/2}$, and the power $P=10$. Note that
the hybrid coding bound in~\eqref{eq:TWRCnew} strictly outperforms
 amplify--forward and noisy network coding for every~$r \in (0,1/2)$.
%Here the gain is due to the fact
%that hybrid coding provides differentiated information (analog/digital) to separate receivers.
%Hybrid coding also outperforms decode--forward when the relay is sufficiently
 %far from both destination nodes.

%
%By setting $U_3 = (\Yh_3, X_3)$ for $p(\yh_3,x_3|y_3) =
%p(\yh_3|y_3)p(x_3)$, this inner bound reduces to the noisy network
%coding inner bound consisting of all rate pairs $(R_1,R_2)$ such that
 %\begin{align*}
%R_1 &< I(X_1; Y_2, \Yh_3|X_2,X_3),\\
%R_1 &< I(X_3; Y_2) - I(Y_3;\Yh_3|X_1,X_2),\\
%R_2 &< I(X_2; Y_1, \Yh_3|X_1,X_3),\\
%R_2 &< I(X_3; Y_1) - I(Y_3;\Yh_3|X_1,X_2)
 %\end{align*}
%for some $p(x_1)p(x_2)p(u_3|y_3)$ and $x_3(u_3,y_3)$.
%

%%%%
\subsubsection{Hybrid Coding Architecture}
\label{sec:2w_hybrid}
The proposed relay coding scheme can be described as follows. A channel encoder at source node~$j=1,2$ maps the message $M_j$ into one of $2^{nR_j}$ sequences $X_j^n(M_j)$ generated i.i.d.\@ according to $\prod_{i=1}^n p_{X_j}(x_{ji})$.
 The relay node uses the hybrid coding architecture introduced in Section~\ref{sec:ptp} for the problem of lossy communication over a point-to-point channel. Specifically, at the relay node, the ``source'' sequence $Y^n_3$ is mapped via hybrid coding to one of $2^{nR_3}$ independently generated sequences $U_3^n(L_3)$ via joint typicality encoding and then the pair $(Y_3^n,U_3^n(L_3) )$ is mapped to $X_3^n$ via the symbol-by-symbol map $x_3(u_3,y_3)$.  Decoding at node $1$  is performed by searching for the unique message $\Mh_2 \in [1:2^{nR_2}]$ such that the tuple $( X^n_1(M_1), U_3^n(L_3),X^n_2(\Mh_2),Y_4^n)$ is jointly typical for some  $L_3 \in [1:2^{nR_3}]$. In other words, node 1 nonuniquely decodes the sequence $U_3^n(L_3)$ selected by the relay node.

The conditions under which a rate pair $(R_1,R_2)$ is achievable can be obtained by studying the conditions for channel decoding at the destinations and for hybrid encoding at the relay separately. By the covering lemma, the encoding operation at the relay node is successful if
 \begin{align*}
R_3> I(Y_3; U_3).
\end{align*}
On the other hand, by the packing lemma, suitably modified to account for the dependence between the index and the codebook at the relay node, the channel decoding operation at node~1 is successful if
 \begin{align*}
R_2 &<  I(X_2;Y_1, U_3|X_1), \\
R_2 + R_3 & < I(X_2,U_3; X_1,Y_1) + I(X_2;U_3).
%R_2 &< \min \bigl( I(X_2;Y_1, U_3|X_1), I(X_2,U_3; X_1,Y_1) - I(Y_3;U_3|X_1
\end{align*}
 Similar conditions hold for the case of decoder 2. The lower bound~\eqref{eq:dm-twrc} is then established by combining the above inequalities and eliminating the intermediate rate $R_3$.

%Notice that in this scheme the channel output $Y_3$ is mapped by the relay node to a single codeword $U$ which is decoded by both destination nodes by indirect decoding. A more general scheme similar to the one in Theorem~\ref{thm:ic} would allow a mapping from $Y_3$ to three auxiliary random variables, one of which is decoded by both destination nodes while the remaining two are decoded by only one of the two receivers. Despite its simplicity, however, Theorem~\ref{thm:dm-twrc} can strictly outperform many of the existing coding schemes, as demonstrated in the following special case.

\subsection{Diamond Relay Network}\label{sec:diamond}

\begin{figure}[h]
\centering
\footnotesize
\psfrag{m}[b]{$M$}
\psfrag{mh}[b]{$\Mh$}
\psfrag{x}[b]{$X_1^n$}
\psfrag{p}[c]{$p(y_2,y_3|x_1)$}
\psfrag{p2}[c]{$p(y_4|x_2,x_3)$}
\psfrag{y}[b]{$Y_4^n$}
 \psfrag{y1}[b]{$Y_2^n$}
\psfrag{y2}[b]{$Y_3^n$}
\psfrag{m1}[b]{$X_2^n$}
\psfrag{m2}[b]{$X_3^n$}
\psfrag{e}[c]{1}
\psfrag{d}[c]{4}
\psfrag{d1}[c]{2}
\psfrag{d2}[c]{3}
\includegraphics[scale=0.5]{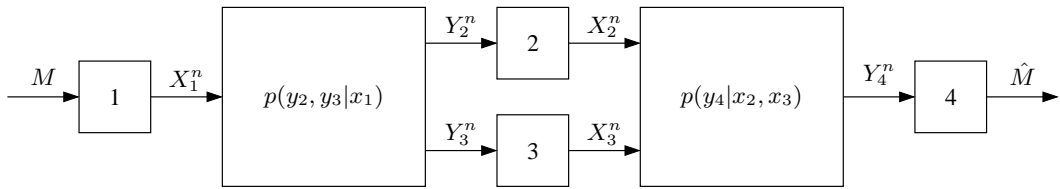}
 \caption{Diamond channel.}
\label{fig:diamond}
\end{figure}

A canonical channel model used to feature the benefits of node cooperation in relay networks is the diamond channel  introduced in~\cite{schein--gallager2000}; see Fig.~\ref{fig:diamond}.
This two-hop network consists of a source node (node 1) that wishes to send a message $M \in [1:2^{nR}]$ to a destination (node~4) with the help of two relay nodes (nodes 2 and 3). The source node is connected through the broadcast channel $p(y_2,y_3|x_1)$ to the two relay nodes that are in turn connected to the destination node through the multiple-access channel $p(y_4|x_2,x_3)$.  A diamond channel $(\Xc_1 \times \Xc_2 \times \Xc_3, p(y_2,y_3|x_1)p(y_4|x_2,x_3),\Yc_2 \times \Yc_3 \times \Yc_4 )$ consists of six alphabet sets and a collection of conditional pmfs on $\Yc_2 \times \Yc_3 \times \Yc_4 $. A $(2^{nR},n)$ code for the diamond channel consists of
 \begin{itemize}
\item a message set $[1:2^{nR}]$,
\item an encoder that assigns a codeword  $x^n_1(m)$ to each message $m\in [1:2^{nR}]$,
\item two relay encoders, where relay encoder $j=2,3$ assigns a symbol $x_{j,i}( y_j^{i-1})$ to each past received output sequence $y_j^{i-1} \in \Yc_j^{i-1}$, and
 \item a decoder that assigns an estimate $\mh$ or an error message to each received sequence $y_4^n \in \Yc_4^n$.
\end{itemize}
We assume that the message $M$ is uniformly distributed over
$[1:2^{nR}]$. The average probability of error is defined as $P_e^{(n)} = \P\{ \Mh \ne M \}$. A rate $R$ is said to be achievable for the diamond channel if there exists a sequence of $(2^{nR},n)$ codes such that $\lim_{n \to \infty} P_e^{(n)} = 0$. The capacity~$C$ of the diamond channel is the supremum of the achievable rates $R$.

The capacity of the diamond channel is not known in general. Schein and Gallager\cite{schein--gallager2000} characterized inner bounds on the capacity region based on decode--forward,  compress--forward, and amplify--forward.
 %As for the DM-TWRC, most of the existing inner bounds~\cite{schein--gallager2000} are based on coding techniques originally developed for the relay channel.
%Such techniques sometimes yield the capacity region~\cite{kang--ulukus} or an approximation of it~\cite{urs--diggavi} for special classes of channels.
 %As before, however, we argue that it is natural to follow a joint source--channel coding approach. Indeed,
%if the codeword $X_1^n(M)$ broadcasted by the source node is an i.i.d.\@ copies of $X_1 \sim p(x_1)$, then the channel outputs at the two relay nodes are i.i.d.\@ copies of the 2-DMS $(Y_2,Y_3) \sim \sum_{x_1}p(x_1)p(y_2,y_3|x_1)$ and thus the hybrid coding scheme presented in Section~\ref{sec:networks} can be used to communicate the source pair $(Y_2,Y_3)$ over the DM-MAC $p(y_4|x_2,x_3)$.
 %

Hybrid coding yields the following lower bound on the capacity, the proof
of which is given in
Appendix~\ref{app:diamond}.
%
%\begin{figure}[h]
%\centering
%\footnotesize
%\psfrag{s}[b]{$Y_k^n$}
 %\psfrag{m}[b]{$U_k^n(L_k)$}
%\psfrag{e1}[c]{\footnotesize Source}
%\psfrag{e2}[c]{\footnotesize Channel}
%\psfrag{d1}[c]{\footnotesize Source}
%\psfrag{d2}[c]{\footnotesize Channel}
%\psfrag{e3}[c]{$x_k(u_k,y_k)$}
 %\psfrag{d3}[c]{$\sh(u,y)$}
%\psfrag{e}[c]{\footnotesize encoder}
%\psfrag{p}[c]{$p(y|x)$}
%\psfrag{d}[c]{\footnotesize decoder}
%\psfrag{x}[b]{$X_k^n$}
%\psfrag{y}[B]{$Y^n$}
%\psfrag{mh}[b]{$U^n(\Mh)$}
 %\psfrag{sh}[B]{$\hat{S}^n$}
%\includegraphics[scale=0.5]{fig/relay.eps}
%\caption{Hybrid coding interface for relays.}
%\label{fig:relay}
%\end{figure}

\begin{theorem} \label{thm:diamond}
The capacity of the diamond channel $p(y_2,y_3|x_1)p(y_4|x_2,x_3)$
 is lower bounded as
\begin{align} \label{eq:diam1}
C \ge \max \min \{ & I(X_1; U_2, U_3, Y_4),  I(X_1,U_2; U_3, Y_4)  - I(U_2; Y_2|X_1),  \nonumber \\
           & I(X_1,U_3; U_2, Y_4) - I(U_3; Y_3|X_1),   I(X_1,U_2, U_3; Y_4) - I(U_2, U_3;Y_2,Y_3|X_1)  \},
 \end{align}
where the maximum is over all conditional pmfs $p(x_1)p(u_2|y_2)p(u_3|y_3)$ and functions
$x_2(u_2,y_2)$, $x_3(u_3,y_3)$.
\end{theorem}

\medskip

%%%
\begin{remark}
Theorem~\ref{thm:diamond} includes both the noisy network coding inner bound, which is recovered by setting $U_j = (X_j,\Yh_j)$ with $p(x_j)p(\yh_j|y_j)$, $j=2,3$, and the amplify--forward inner bound, which is obtained by setting $U_j = \emptyset$ for $j=2,3$, and the inclusion can be strict in general, as demonstrated below.
 \end{remark}

%\subsubsection{Deterministic Diamond Relay Network}\label{sec:detdiamond}
\subsubsection{Deterministic Diamond Channel}
Consider the special case where the multiple access channel
$p(y_2,y_3|x_1)$ and the broadcast channel $p(y_4|x_2,x_3)$ are
 deterministic, i.e., the channel outputs are functions of the corresponding inputs. In this case,
Theorem~\ref{thm:diamond} simplifies to the following.
\medskip
\begin{corollary}
\label{cor:det-diamond}
The capacity of the deterministic diamond channel is lower bounded as
 \begin{align}
C \ge \max_{p(x_1)p(x_2|y_2)p(x_3|y_3)} R(Y_2,Y_3,Y_4|X_2,X_3),
\label{eq:dia1}
\end{align}
where
\begin{align*}
R(Y_2,Y_3,Y_4|X_2,X_3) = \min\{ & H(Y_2,Y_3),
H(Y_2) + H(Y_4|X_2,Y_2), \\
 & \; H(Y_3) + H(Y_4|X_3,Y_3),H(Y_4)\}.
\end{align*}
\end{corollary}
\medskip
\begin{IEEEproof}
Set in Theorem~\ref{thm:diamond} $U_2=(Y_2,X_2)$, $U_3=(Y_3,X_3)$, $x_2(u_2,y_2)=x_2$, and $x_3(u_3,y_3)=x_3$ under a pmf $p(x_2|y_2)p(x_3|y_3)$.
\end{IEEEproof}

We can compare the result in Corollary~\ref{cor:det-diamond} with the existing inner and outer bounds for this channel model. An outer bound on the capacity region is given by the cutset bound~\cite{El-Gamal1981b}, which in this case simplifies to
 \begin{equation}
\label{eq:dia_cb}
C \le \max_{p(x_1)p(x_2,x_3)} R(Y_2,Y_3,Y_4|X_2,X_3)
\end{equation}
On the other hand, specializing the scheme in~\cite{Avestimehr--Diggavi--Tse2009} for deterministic relay networks, we obtain the lower bound
 \begin{equation}
\label{eq:dia_av}
C \ge \max_{p(x_1)p(x_2)p(x_3)} R(Y_2,Y_3,Y_4|X_2,X_3).
\end{equation}
Note that~\eqref{eq:dia1},~\eqref{eq:dia_cb}, and~\eqref{eq:dia_av} differ only
in the set of allowed maximizing input pmfs. In particular,~\eqref{eq:dia1} improves upon the inner bound~\eqref{eq:dia_av}
 by allowing $X_2$ and $X_3$ to
depend on $Y_2$ and $Y_3$ and thereby increasing the set of distributions $p(x_2, x_3)$.  The following example demonstrates that the inclusion can be
strict.

\begin{example}
 \label{example1}
Suppose that $p(y_2,y_3|x_1)$ is the Blackwell broadcast channel
(i.e., $X_1 \in \{0,1,2\}$ and $p_{Y_2,Y_3|X_1}(0,0|0) = p_{Y_2,Y_3|X_1}(0,1|1) =
p_{Y_2,Y_3|X_1}(1|2) = 1$) and $p(y_4|x_2,x_3)$ is the binary
 erasure multiple access channel (i.e., $X_2, X_3 \in \{0,1\}$ and $Y_4 = X_2 + X_3 \in
\{0,1,2\}$). It can be easily seen that the general lower bound
reduces to $C \ge 1.5$, while the capacity is $C = \log 3$, which
 coincides with the hybrid coding lower bound (with $X_2 = Y_2$ and
$X_3 = Y_3$). Thus, hybrid coding strictly outperforms the coding
scheme by Avestimehr, Diggavi, and
Tse~\cite{Avestimehr--Diggavi--Tse2009} and noisy network
 coding~\cite{Lim--Kim--El-Gamal--Chung2011}.
\end{example}
\subsubsection{Hybrid Coding Architecture}
\label{sec:dia_hybrid}

The proof of achievability of Theorem~\ref{thm:diamond} is based on a hybrid coding architecture similar to the one used in the proof of Theorem~\ref{thm:mac} and can be described as follows. At the source node, the message $M$ is mapped to one of $2^{nR_1}$ sequences $X_1^n(M)$ i.i.d. $\sim p(x_1)$ as in point-to-point communication. At the relay nodes, the ``source'' the sequence $Y^n_j$, $j=2,3$, is separately mapped into one of $2^{nR_j}$ independently generated sequences $U_j^n(M_j)$. Then, the pair $(Y_j^n,U_j^n(M_j))$ is mapped by node $j$ to $X_j^n$ via a symbol-by-symbol map. By the covering lemma, the source encoding operation at the relays is successful if
 \begin{align*}
R_2 & > I(U_2;Y_2) \\
R_3 & > I(U_3 ; Y_3).
\end{align*}
At the destination node, decoding is performed by joint typicality and indirect decoding of the sequences $(U_2^n,U_3^n)$, that is, by searching for the unique message $\Mh \in [1:2^{nR}]$ such that the tuple $( X^n_1(\Mh), U_2^n(M_2) , U_3^n(M_3) , Y_4^n  )$ is typical for some  $M_2 \in [1:2^{nR_2}]$ and $M_3 \in [1:2^{nR_3}]$. By the packing lemma, combined with the technique introduced in Section~\ref{sec:ptp}, the channel decoding operation at the destination node is successful if
 \begin{align*}
R & < I(X_1 ; U_2,U_3,Y_4 ) \\
R + R_2 & < I(X_1, U_2; U_3, Y_4)+I(X_1;U_2) \\
R + R_3 & < I(X_1, U_3; U_2, Y_4)+I(X_1;U_3) \\
R + R_2 + R_3 & < I(X_1, U_2, U_3; Y_4)+I(X_1;U_2)+I(X_1,U_2;U_3).
 \end{align*}
Hence, the lower bound~\eqref{eq:diam1} is obtained by combining the conditions for source coding at the relay nodes with those for channel decoding at the destination and by eliminating the auxiliary rates $(R_1,R_2)$ from the resulting system of inequalities.

%\subsection{Discussion}
%\label{sec:relay_dis}
%
%In this section we presented two channel coding schemes for relay networks in which each relay uses the hybrid coding interface introduced in Section~\ref{sec:ptp} to forward the received channel output to the destination.

\section{Concluding Remarks}
\label{sec:con}

In this paper we presented a new approach to studying lossy communication of correlated sources over networks based on hybrid analog/digital coding.  We first revisited the problem of lossy communication over a point-to-point channel, for which we proposed a hybrid scheme that generalizes both the digital, separate source and channel coding scheme and the analog, uncoded transmission scheme. Similar to Shannon's source--channel separation architecture, the proposed hybrid scheme employs a modular system architecture, whereby the source encoding and channel decoding operations are decoupled. However, unlike the separation architecture, a single codebook is used for both source coding and channel coding, which
 allows the resulting coding scheme to achieve the performance of the
best known joint source--channel coding schemes.

Next, we discussed how the proposed hybrid coding architecture can be generalized to construct joint source--channel coding schemes for lossy communication over multiuser channels. To illustrate the main ideas, we focused on the specific problem of lossy communications over multiple access channels, for which we presented a joint source--channel coding scheme that unifies and generalizes several existing results in the literature. As in the case of point-to-point communication, the proposed scheme is conceptually similar to separate source and channel coding and is obtained by concatenating the source coding scheme in~\cite{Berger1978,Tung1978} for distributed lossy source coding with a channel code for multiple access communication, except that the same codeword is used for source coding as well as for channel coding. The same design principle can be readily adapted to other joint source--channel coding problems for which separate source coding and channel coding have matching index structures, such as
 \begin{itemize}
\item communication of a 2-DMS with common part over a DM-MAC
  (Berger--Tung coding with common part~\cite{Kaspi--Berger1982,
  Wagner--Kelly--Altug2009} matched to the multiple access channel coding
   with common message~\cite{Slepian--Wolf1973b}),

\item communication of a 2-DMS over a DM broadcast channel (lossy Gray--Wyner
  system~\cite{Gray--Wyner1974} matched to Marton's coding for a
  broadcast channel~\cite{Marton1979}),
 %\item communication of a bivariate Gaussian source over a Gaussian
%  broadcast channel~\cite{Tian--Diggavi--Shamai2010}, and
\item communication of a 2-DMS over a DM interference channel (extension of Berger--Tung
   coding for a 2-by-2 source network matched to Han--Kobayashi coding
  for an interference channel~\cite{Han--Kobayashi1981}).
\end{itemize}
In all these cases, hybrid coding performs as well as (and sometimes better than) the existing coding schemes~\cite{FaT}.

%As a motivation, in such networks the signal transmitted by the source node results in correlated signals received at the intermediate relays, which can then cooperatively forward the transmitted information to the destination node using joint source--channel coding techniques~\cite{Kochman-ela2008,Yao--Skoglund2009,Khormuji--Skoglund2010,Khormuji--Skoglund2011}.
 %~\cite{Kochman-ela2008,Yao--Skoglund2009,Khormuji--Skoglund2010,Khormuji--Skoglund2011}

Finally, we explored applications of hybrid coding in the context of relay networks. %Joint source--channel coding techniques were previously used in the literature to design coding scheme for specific Gaussian relay networks:~\cite{Yao--Skoglund2009} considers a Gaussian relay channel with flat fading and no channel state information at the transmitter and uses hybrid digital--analog coding as technique to increase the robustness against the unknown fading in the transmission from the relay to the destination;~\cite{Kochman-ela2008} considers a Gaussian parallel relay network with colored noise wherein the source-to-relay and relay-to-destination links have mismatched bandwidths, and uses joint source--channel coding techniques to compress the received signal at the relay and forward it to the destination.
 We introduced a general coding technique for DM relay networks based on hybrid coding, whereby each relay uses the hybrid coding interface to transmit a symbol-by-symbol function of the received sequence and its quantized version (analog-to-analog/digital interface). We demonstrated via two specific examples, the
 two-relay diamond channel and the two--way relay channel, that the proposed hybrid coding can strictly outperform both amplify--forward (analog-to-analog interface) and compress--forward/noisy network coding (analog-to-digital interfaces). For simplicity, we assumed that the relay nodes do not attempt to decode the message transmitted by the source, but the presented results can be further improved by combining hybrid coding with other coding techniques such as decode--forward and structured coding~\cite{Nazer--Gastpar2007}. In this case, hybrid coding provides a general analog/digital-to-analog/digital interface for relay communication.
 While we have focused on two specific examples, similar ideas can be applied to general layered network model, provided that the proposed hybrid coding scheme is repeated at each layer in the network~\cite{FaT}. In principle, hybrid coding can also be applied to the relay channel and other nonlayered relay networks. However, in this case hybrid coding (or even amplify--forward) would not yield inner bounds to the capacity region in a single-letter form, due to the dependency between the channel input at each relay node and the previously received analog channel outputs.

%Joint source--channel coding techniques have been previously used in~\cite{Kochman-ela2008,Yao--Skoglund2009,Khormuji--Skoglund2010,Khormuji--Skoglund2011} to design coding scheme for specific Gaussian relay networks.

%%%%%%%%%%%%%%%%%%

%%%%%%%%%%%%%%%%%%%%%%%%%%%%%%%%%%%%%%%%%%%%%%%
\appendices
%%%%%%%%%%%%%%%%%%%%%%%%%%%%Appendix Proof of
\section{Proof of Lemma~\ref{lem:1}}
\label{app:1}
Given $\ut^n$ and $s^n$,
 let $\Ac = \{ U^n(1)=\ut^n, S^n=s^n\}$ in short. Let $\Cc'= \{ U^n(m): m\in[3:2^{nR}] \}$. Then, by the law of total probability and the Bayes rule, for every $u^n$,
\begin{align}
\P&\{U^n(2)=u^n \cond M=1,\Ac\} \nonumber \\
 &=\sum_{\tiny\Cc'} \P\{\Cc'=\cc',U^n(2)=u^n \cond M=1,\Ac \} \nonumber \\
&=\sum_{\tiny\Cc'} \P\{\Cc'=\cc' \cond M=1,\Ac\}\, \P\{U^n(2)=u^n \cond \Ac, \Cc'=\cc'\} \,
     \frac{\P\{M=1 \cond \Ac, U^n(2)=u^n, \Cc'=\cc' \}}{\P\{M=1 \cond \Ac, \Cc'=\cc'\}} \nonumber \\
 &= \sum_{\tiny\Cc'}  \P\{\Cc'=\cc' \cond M=1, \Ac\}\, \biggl(\prod_{i=1}^n p_{U}(u_i) \biggr) \,
     \frac{\P\{M=1\cond \Ac, U^n(2)=u^n, \Cc'=\cc' \}}{\P\{M=1 \cond \Ac, \Cc'=\cc'\}}. \label{eq:lem1-0}
 \end{align}
%where the summation is throughout over $\cc' = \{u^n(m): m \in [3:2^{nR}]\}$ such that  $\P\{\Cc'=\cc' \cond M=1,\Ac\}>0$. Next, we bound the last term on the right hand side of~\eqref{eq:lem1-0}.
 For each $(\ut^n,s^n,u^n,\cc')$ such that $\P\{\Cc'=\cc'|M=1,\Ac\}>0$, let
$n(\ut^n,s^n,u^n, \cc') = n(s^n, \cc') = | \{ u'^n \in \cc': (u'^n, s^n ) \in \aepvar \} |$
denote the number of unique sequences in $\cc'$ %(up to multiplicity) 
that are jointly typical with $s^n$ and
 \begin{equation*}
i(\ut^n, s^n, u^n, \cc') = i(\ut^n,s^n,\cc')=
\begin{cases}
1, &  (\ut^n,s^n ) \not \in \aepvar \text{ and } n(s^n,\cc') = 0,\\
0, & \text{otherwise},
\end{cases}
\end{equation*}
 be the indicator function for the case that neither $\ut^n$ nor any codeword in $\cc'$ is jointly typical with $s^n$. Then, by the way the random index $M$ is generated, it can be easily verified that
\begin{equation*}
 \P\{M=1 \cond \Ac,  U^n(2)= u^n, \Cc'=\cc' \} \le
\frac{1}{2^{nR}}i(\ut^n,s^n,\cc')  + \frac{1}{n(s^n,\cc') + 1} (1- i(\ut^n,s^n,\cc')). \label{eq:lem1-1}
\end{equation*}

Similarly, since $U^n(2) \sim \prod_{i=1}^n p_U(u_i)$, independent of $S^n$ and $U^n(m)$, $m\ne 2$,
 \begin{align*}
\P&\{M=1 \cond  \Ac , \Cc'=\cc \} \nonumber\\
&\ge \P\{M=1 \cond  \Ac , \Cc'=\cc', 2 \not \in \Ic \} \cdot \P\{ 2 \not \in \Ic \cond \Ac, \Cc'=\cc' \}  \nonumber \\
&\ge \P\{M=1 \cond  \Ac , \Cc'=\cc', 2 \not \in \Ic \}  \left(1-2^{-n(I(U;S)-\d(\e'))}\right) \nonumber  \\
 &= \left( \frac{1}{2^{nR}}i(\ut^n,s^n,\cc')  + \frac{1}{n(s^n,\cc') + 1} (1- i(\ut^n,s^n,\cc')) \right)
\left(1-2^{-n(I(U;S)-\d(\e'))}\right) \label{eq:lem1-2}.
\end{align*}
It follows that
 \begin{equation}
    \frac{\P\{M=1|  U^n(2)=u^n,E,\Cc'=\cc' \}}{\P\{M=1| E, \Cc'=\cc'\}}\le \frac{1}{1-2^{-n(I(U;S)-\d(\e'))}} \le 1 + \e \label{eq:lem1-3}
\end{equation}
for $n$ sufficiently large. By combining~\eqref{eq:lem1-0} and~\eqref{eq:lem1-3}, the claim follows.

\section{Proof of Theorem~\ref{thm:mac}}
\label{sec:mac_achie}

For simplicity, we consider the case $Q = \emptyset$. Achievability for an arbitrary $Q$ can be proved
using coded time sharing~\cite[Section~4.5.3]{El-Gamal--Kim2011}.

\medskip
\emph{Codebook generation:} Let $\e>\e'>0$. Fix a pmf $p(u_1|s_1)p(u_2|s_2)$, channel encoding functions $x_1(u_1,s_1)$ and $x_2(u_2,s_2)$, and source decoding functions $\sh_1(u_1,u_2,y)$ and
 $\sh_2(u_1,u_2,y)$ such that $\E( d_j(S_j,\Sh_j) ) \le D_j/(1+\e)$, $j=1,2$. For each $j = 1, 2$, randomly and independently generate $2^{nR_{j}}$ sequences $u^n_j(m_j)$, $m_j\in[1:2^{nR_{j}}]$, each according to $\prod_{i=1}^n p_{U_j}(u_{ji})$. The codebook $\Cc = \{  (u_1^n(m_1),
 u_2^n(m_2)): m_1\in[1:2^{nR_1}] \times [1:2^{nR_2}] \}$ is revealed to both the encoders and the decoder.

\medskip
\emph{Encoding:} Upon observing a sequence $s_j^n$,
encoder $j = 1,2$ finds an index $m_j \in [1:2^{nR_j}]$ such that $(s_j^n, u_j^n(m_j) ) \in \aepvar$.
 If there is more than one such
index, it chooses one of them at random.  If there is no such index,
it chooses an arbitrary index at random from $[1:2^{nR_j}]$.  Encoder
$j$ then transmits $x_{ji} = x_j( u_{ji}(m_j) , s_{ji})$ for $i \in [1:n]$.

\emph{Decoding:} Upon receiving $y^n$, the decoder finds the
unique index pair $(\mh_1, \mh_2)$ such that
$(u_1^n(\mh_1),u_2^n(\mh_2), y^n) \in \aep$
and sets the estimates as
$\sh_{ji} = \sh_j( u_{1i}(m_1),u_{2i}(m_2),y_i)$, $i \in [1:n]$, for
 $j = 1,2$.

\emph{Analysis of the expected distortion:} We
bound the distortion averaged over $(S_1^n, S_2^n)$, the random choice of
the codebook $\Cc$, and the random index assignments at the encoders. Let $M_1$ and $M_2$ be the random variables denoting the chosen
 indexes at encoder 1 and at encoder 2, respectively. Define the ``error'' event
\[
\Ec=\bigl\{(S_1^n, S_2^n, U_1^n(\Mh_1), U_2^n(\Mh_2), Y^n) \not\in \aep \bigr\}
\]
and partition it into
\begin{align*}
 \Ec_j &= \bigl\{(S_j^n, U_j^n(m_j) ) \not\in \aepvar \text{ for all } m_j \bigr\}, \quad j=1,2,\\
\Ec_3 &= \bigl\{ (S_1^n, S_2^n, U_1^n(M_1), U_2^n(M_2), Y^n)  \not\in \aep \bigr\},\\
\Ec_4 &= \bigl\{(U_1^n(m_1), U_2^n(m_2),  Y^n) \in \aep  \text{ for some } m_1 \ne M_1, m_2\ne M_2 \bigr\}, \\
 \Ec_5 &= \bigl\{(U_1^n(m_1), U_2^n(M_2),  Y^n) \in \aep \text{ for some } m_1\ne M_1 \bigr\},\\
\Ec_6 &= \bigl\{(U_1^n(M_1), U_2^n(m_2),  Y^n) \in \aep \text{ for some } m_2\ne M_2 \bigr\}.
\end{align*}
Then by the union of events bound,
 \begin{align*}
\P(\Ec) &\le \P(\Ec_1 )+\P(\Ec_2 )+\P(\Ec_3  \cap \Ec_1^c\cap \Ec_2^c)+\P(\Ec_4)+\P(\Ec_5)+\P(\Ec_6).
\end{align*}
As in the case of point-to-point communication studied in Section~\ref{sec:ptp}, the desired distortion pair is achieved if $\P(\Ec)$ tends
 to zero as $n\to \infty$.  By the covering lemma, $\P(\Ec_1 )$ and
$\P(\Ec_2 )$ tend to zero as $n\to\infty$, if
\begin{align}
R_1 &> I(U_1; S_1)+\d(\e'), \label{eq:mac1}\\
R_2 &> I(U_2; S_2)+\d(\e'). \label{eq:mac2}
 \end{align}
By the Markov lemma~\cite[Section~12.1.1]{El-Gamal--Kim2011}, the
third term tends to zero as $n\to \infty$.

By the symmetry of random codebook generation and encoding, we analyze the remaining probabilities conditioned on the event  $\Mc = \{M_1 = 1, M_2 = 1\}.$ First, we bound $\P(\Ec_4)$. By the union of events bound,
 \begin{align}
\P (\Ec_4 | \Mc )
     \le   \sum_{m_1 =2}^{2^{nR_1}}  \sum_{m_2 =2}^{2^{nR_2}} \sum_{ (u_1^n,u_2^n,y^n) \in \aep}
                                  \P \bigl\{ U_1^n(m_1)=u_1^n, U_2^n(m_2)=u_2^n, Y^n=y^n |\Mc \bigr\}.
 \label{eq:MAC_0}
\end{align}
Let $\Ut^n = ( U_1^n(1), U_2^n(1), S_1^n, S_2^n  )$
and $\ut^n = (\ut_1^n, \ut_2^n, s_1^n, s_2^n)$ in short.
Then, by the law of total probability, for $m_1 \ne 1$ and $m_2 \ne 1$,
 \begin{align}
\P& \bigl\{ U_1^n(m_1)=u_1^n, U_2^n(m_2)=u_2^n, Y^n=y^n |\Mc \bigr\} \nn \\
&= \sum_{ \ut^n }  \P \bigl\{ U_1^n(m_1)=u_1^n, U_2^n(m_2)=u_2^n, Y^n=y^n, \Ut^n=\ut^n |\Mc \bigr\} \nn \\
&\stackrel{(a)}{=}  \sum_{ \ut^n}  \P \bigl\{ U_1^n(m_1)=u_1^n \cond \Mc , \Ut^n=\ut^n  \bigr\}
                                    \P \bigl\{ U_2^n(m_2)=u_2^n | \Mc, U_1^n(m_1)=u_1^n, \Ut^n=\ut^n \bigr\}    \nn \\[-1em]
& \quad \qquad \times  \P \{ \Ut^n=\ut^n  \cond  \Mc , Y^n=y^n \} \P\{ Y^n=y^n | \Mc \}  \nn \\
&\stackrel{(b)}{=}  \sum_{(\ut_1^n,\ut_2^n, s_1^n,s_2^n)}  \P \bigl\{ U_1^n(m_1)=u_1^n \cond M_1 = 1, U_1^n(1)=\ut_1^n, S_1^n=s_1^n  \bigr\}  \nn \\[-1em]
&\qquad\qquad \qquad \times  \P \bigl\{ U_2^n(m_2)=u_2^n | M_2 = 1, U_2^n(1) = \ut_2^n, S_2^n = s_2^n \bigr\}    \nn \\
& \qquad\qquad \qquad \times  \P \{ \Ut^n=\ut^n  \cond  \Mc , Y^n=y^n \} \P\{ Y^n=y^n | \Mc \}  \nn \\
&\stackrel{(c)}{\le}  (1 + \e)   \sum_{ \ut^n}  \biggl(\prod_{i=1}^np_{U_1}( u_{1i})  p_{U_2}( u_{2i})\biggr)  \P\{ \Ut^n=\ut^n  |  \Mc , Y^n=y^n \}   \P\{Y^n= y^n |\Mc\} \nn  \\
&=  (1 + \e)   \biggl(\prod_{i=1}^np_{U_1}( u_{1i})  p_{U_2}( u_{2i})\biggr)  \P\{Y^n= y^n |\Mc\} \label{eq:MAC_2}
\end{align}
for $n$ sufficiently large.  Here, $(a)$ follows from the fact that given $\Mc$
 \begin{equation}
\label{eq:MC1}
(U_1^n(m_1), U_2^n(m_2)) \to ( S^n_1,S^n_2, M_1,M_2,U_1^n(M_1), U_2^n(M_2)) \to Y^n
\end{equation}
form a Markov chain for all $m_1 \ne 1$ and $m_2 \ne 1$, while $(b)$ follows by the independence of the sequences
 and the encoding procedure. For step $(c)$, we
apply Lemma~\ref{lem:1} twice.
Combining~\eqref{eq:MAC_0} and~\eqref{eq:MAC_2}, it follows that for $n$ sufficiently large
\begin{align}
\P (\Ec_4 | \Mc )
& \le  (1+\e)   \sum_{m_1 =2}^{2^{nR_1}}   \sum_{m_2 =2}^{2^{nR_2}}  \sum_{(u_1^n,u^n_2, y^n) \in \aep}  p_{U^n_1}( u_1^n ) p_{U^n_2}( u_2^n )  \P\{Y^n= y^n |\Mc\} \nn  \\
&  \le  (1+\e)  \; 2^{n(R_1+R_2)}  \sum_{y^n \in \aep}  \P\{ Y^n=y^n|\Mc\}  2^{-n (I(U_1,U_2;Y) + I(U_1;U_2) -\d( \epsilon) ) }  \nn \\
&  \le  (1+\e) \; 2^{n (R_1+R_2 - I(U_1,U_2;Y) - I(U_1;U_2) + \d(\epsilon) )   }.\nn
 \end{align}
Hence, $\P(\Ec_4)$ tends to zero as $n \to \infty$ if
\begin{equation}
\label{eq:mac3}
R_1+R_2 < I(U_1,U_2;Y) + I(U_1;U_2) -\d(\epsilon).
\end{equation}
Following similar steps, $\P(\Ec_5)$ is upper bounded by
 \begin{align}
\P&\bigl\{  (U_1^n(m_1), U_2^n(1), Y^n) \in \aep \text{ for some }  m_1 \ne 1 |\Mc \bigr \} \nn \\
& \le \sum_{m_1 =2}^{2^{nR_1}} \P\bigl\{ (U_1^n(m_1), U_2^n(1), Y^n) \in \aep |\Mc   \bigr\} \nn \\
& =   \sum_{m_1 =2}^{2^{nR_1}} \sum_{ (u_1^n,u_2^n,y^n) \in \aep}  \P\bigl\{ U_1^n(m_1)=u_1^n, U_2^n(1)=u_2^n, Y^n=y^n |\Mc \bigr\}   \nn \\
&\stackrel{(a)}{\le}  (1+\e)   \sum_{m_1 =2}^{2^{nR_1}}  \sum_{(u_1^n,u^n_2, y^n) \in \aep}  p_{U^n_1}( u_1^n ) \P\bigl\{Y^n= y^n ,   U^n_2(1) = u_2^n |\Mc \bigr\} \nn \\
&  \le (1+\e) 2^{n R_1 }  \sum_{(u_2^n,y^n) \in \aep} \P\bigl\{ U_2^n(1)=u_2^n ,Y^n= y^n |\Mc\bigr\} 2^{-n (I(U_1;Y,U_2) -\d( \epsilon) ) }  \nn \\
&  \le (1+\e) 2^{n (R_1-I(U_1;Y,U_2) +\d( \epsilon) ) }  \nn
 \end{align}
for $n$ sufficiently large, which implies that $\P(\Ec_5)$ tends to zero as $n \to \infty$ if
\begin{equation}
\label{eq:mac4}
R_1 < I(U_1;Y,U_2) -\d(\epsilon).
\end{equation}
%%%%%%%%%%%%%%%%%%%%%%%%%%%%%%
 %%%%%%%%%%%%%%%%%%%%%%%%%%%%%%
In the above chain of inequalities step (a) is justified as follows. Let $\Ut^n = ( U_1^n(1), S_1^n, S_2^n  )$
and $\ut^n = (\ut_1^n, s_1^n, s_2^n)$ in short.
Then, by the law of total probability, for $m_1 \ne 1$ and $m_2 \ne 1$,
 \begin{align}
\P& \bigl\{ U_1^n(m_1)=u_1^n, U_2^n(1)=u_2^n, Y^n=y^n |\Mc \bigr\} \nn \\
&= \sum_{ \ut^n }  \P \bigl\{ U_1^n(m_1)=u_1^n, U_2^n(1)=u_2^n, Y^n=y^n, \Ut^n=\ut^n |\Mc \bigr\} \nn \\
&\stackrel{(b)}{=}  \sum_{ \ut^n}  \P \bigl\{ U_1^n(m_1)=u_1^n \cond \Mc , U_2^n(1)=u_2^n, \Ut^n=\ut^n  \bigr\}   \P \bigl\{ \Ut^n=\ut^n  \cond  \Mc , U_2^n(1)=u_2^n, Y^n=y^n \bigr\}
                                       \nn \\[-1em]
& \quad \qquad \times  \P\bigl\{ U_2^n(1)=u_2^n, Y^n=y^n | \Mc \bigr\}  \nn \\
&\stackrel{(c)}{=}  \sum_{\ut^n}  \P \bigl\{ U_1^n(m_1)=u_1^n \cond M_1 = 1, U_1^n(1)=\ut_1^n, S_1^n=s_1^n  \bigr\}  \nn \\[-1em]
 &\quad\qquad \times   \P \bigl\{ \Ut^n=\ut^n  \cond  \Mc , U_2^n(1)=u_2^n, Y^n=y^n \bigr\} \P\bigl\{ U_2^n(1)=u_2^n, Y^n=y^n | \Mc \bigr\}    \nn \\
&\stackrel{(d)}{\le}  (1 + \e)   \sum_{ \ut^n}  \biggl(\prod_{i=1}^np_{U_1}( u_{1i})  \biggr)  \P \bigl\{ \Ut^n=\ut^n  \cond  \Mc , U_2^n(1)=u_2^n, Y^n=y^n \bigr\}  \nn \\[-0.5em]
 &\quad\qquad \times\P\bigl\{ U_2^n(1)=u_2^n, Y^n=y^n | \Mc \bigr\} \nn  \\
&=  (1 + \e)   \biggl(\prod_{i=1}^np_{U_1}( u_{1i}) \biggr)  \P\bigl\{ U_2^n(1)=u_2^n, Y^n=y^n | \Mc \bigr\} \nn
\end{align}
 for $n$ sufficiently large.  Here, $(b)$ follows from the fact that~\eqref{eq:MC1} form a Markov chain for all $m_1 \ne 1$ and $m_2 \ne 1$, $(c)$ follows by the independence of the sequences and the encoding procedure, while $(d)$ follows from Lemma~\ref{lem:1}.
 %%%%%%%%%%%%%%%%%%%%%%%%%%%%%%
%%%%%%%%%%%%%%%%%%%%%%%%%%%%%%

Finally, $\P(\Ec_6)$ can be bounded in a similar manner, provided that the subscripts 1 and 2 are
interchanged in the upper bound for $\P(\Ec_5)$. It follows that $\P(\Ec_6)$ tends to zero as $n \to \infty$ if
 \begin{equation}
\label{eq:mac5}
R_2 < I(U_2;Y,U_1) -\d(\epsilon).
\end{equation}
Therefore, if~\eqref{eq:mac1},~\eqref{eq:mac2},~\eqref{eq:mac3},~\eqref{eq:mac4}, and~\eqref{eq:mac5}, the probability of ``error'' tends to zero as
 $n\to\infty$ and the average distortions over the random codebooks is
bounded as desired. Thus, there exists at least one sequence of codes
achieving the desired distortions. By letting $\e \to 0$ and eliminating the intermediate rate pair $(R_1,R_2)$, the sufficient condition in Theorem~\ref{thm:mac} (with $Q = \emptyset$) for lossy communication over a DM-MAC via hybrid coding is established.

\section{Proof of Theorem~\ref{thm:dm-twrc}}
\label{sec:proof-twrc}
We use $b$ transmission blocks, each consisting of $n$ transmissions, as in the proof of the multihop lower bound for the relay channel~\cite[Section~16.4.1]{El-Gamal--Kim2011}. A sequence of $(b-1)$ message pairs $(m_{1j},m_{2j})\in[1:2^{nR_1}]\times [1:2^{nR_2}],$ $j\in[1:b-1]$, each selected independently and uniformly over $[1:2^{nR_1}]\times [1:2^{nR_2}]$ is sent over $b$ blocks. Note that the average rate pair over the $b$ blocks is $({(b-1)}/{b})(R_1,R_2)$, which can be made arbitrarily close to $(R_1,R_2)$ by letting $b\to\infty$.

{\em Codebook generation}: Let $\e > \e'> 0$. Fix $p(x_1)p(x_2)p(u_3|y_3)$ and an encoding function $x_3(u_3,y_3)$. We randomly and independently generate a codebook for each block. For $j\in[1:b]$, randomly and independently generate $2^{nR_3}$ sequences $u_3^n(l_{3j})$, $l_{3j}\in[1:2^{nR_3}],$ each according to $\prod_{i=1}^n p_{U_3}(u_{3i})$. For each $k=1,2,$ randomly and independently generate $2^{nR_k}$ sequences $x^n_k(m_{kj})$, $m_{kj}\in[1:2^{nR_k}]$, each according to $\prod_{i=1}^n p_{X_k}(x_{ki})$.
 This defines the codebook
\[
\Cc_j = \bigl\{ (x_1^n(m_{1j}), x_2^n(m_{2j}), u_3^n(l_{3j}) ):  m_1 \in [1:2^{nR_1}], m_2 \in [1:2^{nR_2}],
l_3 \in [1:2^{nR_3}] \bigr\}
\]
for $j \in [1:b]$.
\medskip

{\em Encoding}: Let $m_{kj}\in[1:2^{nR_k}]$ be the independent message to be sent in block $j\in[1:b-1]$ by node $k=1,2$. Then, node $k$ transmits $x^n_{k}(m_{kj})$ from codebook $\Cc_j$.
\medskip

{\em Relay encoding}: Upon receiving $y^n_3(j)$ in block $j\in[1:b-1]$, relay node 3 finds an index $l_{3j}$ such that $(u^n_k(l_{3j}), y^n_3(j))\in \aepvar$. If there is more than one index, it chooses one of them at random. If there is no such index, it chooses an arbitrary index at random from $[1:2^{nR_3}]$. In block $j+1$, relay $3$ then transmits $x_{3i}= x_{3i}(u_{3i}(l_{3j}), y_{3i}(j))$ for $i\in[1:n]$.
 \medskip

{\em Decoding}: Upon receiving $y^n_1(j)$, $j\in[2:b]$, decoder 1 finds the unique message $\mh_{2,(j-1)}$ such that
\[
    (x^n_1(m_{j-1}), u^n_3(l_{3,(j-1)}), x^n_2(\mh_{2,(j-1)}), y^n_1(j))\in \aep,
 \]
for some $l_{3,(j-1)}\in[1:2^{nR_3}]$. Decoding at node 2 is performed in a similar manner.
\medskip

{\em Analysis of the probability of error}: We analyze the probability of decoding error at node 1 in block $j=2,\ldots,b$, averaged over the random codebooks and index assignment in the encoding procedure at the relay. Let $L_{3,(j-1)}$ be the random variable denoting the index chosen in block $j-1$ at relay 3.  Decoder 1 makes an error only if one or more of the following events occur:
 %The analysis is the same for each transmission block, so we drop the block subscript to simplify notation.  To bound the probability of error, assume without loss of generality that $M=1$.

\begin{align*}
\Ec_1&=\{(Y^n_3(j-1), U^n_3(l))\not\in \aepvar \text{ for all } l \},\\
 \Ec_2&=\{(X^n_1(M_{1,(j-1)}),U^n_3(L_{3,(j-1)}), X^n_2(M_{2,(j-1)}), Y^n_1(j))\not\in \aep\},\\
\Ec_3&=\{(X^n_1(M_{1,(j-1)}), U^n_3(L_{3,(j-1)}), X^n_2(m), Y^n_1(j) ) \in \aep \text{ for some } m\neq M_{2,(j-1)}\},\\
 \Ec_4&=\{(X^n_1(M_{1,(j-1)}), U^n_3(l), X^n_2(m), Y^n_1(j) ) \in \aep \text{ for some } m\neq M_{2,(j-1)} , l \neq L_{3,(j-1)}  \},
\end{align*}
Then, by the union of events bound the probability of decoding error is upper bounded as
 \begin{align*}
\P( \Mh_{2,(j-1)} \neq M_{2,(j-1)}) \le \P(\Ec_1)+\P(\Ec_2\cap \Ec_1^c)+\P(\Ec_3)+\P(\Ec_4).
\end{align*}

By the covering lemma, $\P(\Ec_1)$ tends to zero as $n\to \infty$, if
\begin{align*}
 R_3 &> I(U_3; Y_3)+\d(\e').
\end{align*}
By the Markov lemma, $\P(\Ec_2\cap \Ec_1^c)$ tends to zero as $n\to\infty$.
By the symmetry of the random codebooks generation and random index assignment at the relays, it suffices to consider the conditional probabilities of the remaining error events conditioned on the event that
 \begin{align}
\Mc = \{ M_{1,(j-1)} = 1, M_{2,(j-1)} = 1, L_{3,(j-1)} = 1\}.
\end{align}
Then, by the packing lemma, $\P(\Ec_3)$ tends to zero as $n\to \infty$ if
\begin{equation*}
R_2 < I(X_2; Y_1, U_3| X_1)-\d(\e).
 \end{equation*}
Next,  for $n$ large enough $\P(\Ec_4)$ is upper bounded by
\begin{align}
\P& \bigl\{( X^n_1(1), U^n_3(l), X^n_2(m), Y^n_1(j) ) \in \aep \text{ for some } l \neq 1, m\neq 1 | \Mc \bigr\} \nonumber \\
&\le \sum_{m=2}^{2^{nR_2}}\sum_{l=2}^{2^{nR_3}}  \sum_{(x_1^n,u^n_3,x^n_2,y^n_1) \in \aep} \P\bigl\{X^n_1(1)=x_1^n, U^n_3(l)=u^n_3,X^n_2(m)=x^n_2, Y^n_1(j)=y^n_1| \Mc \bigr\} \nonumber \\
&\stackrel{(a)}{\le}  (1+\e) \sum_{m=2}^{2^{nR_2}}\sum_{l=2}^{2^{nR_3}}  \sum_{(x_1^n,u^n_3,x^n_2,y^n_1) \in \aep} p_{U^n_3}(u^n_3) \cdot \P\bigl\{X^n_1(1)=x_1^n,X^n_2(m)=x^n_2, Y^n_1(j)=y^n_1 | \Mc \bigr\}  \nn \\
 &= (1+\e) \sum_{m=2}^{2^{nR_2}}\sum_{l_2=2}^{2^{nR_2}}  \sum_{(x_1^n,u^n_3,x^n_2,y^n_1) \in \aep} p_{U^n_3}(u^n_3) \cdot p_{X_2^n}(x_2^n)  \cdot \P\bigl\{X^n_1(1)=x_1^n, Y^n_1(j)=y^n_1 | \Mc \bigr\} \nonumber \\
&  \le (1+\e) 2^{n (R_2 + R_3) }  \sum_{(x_1^n,y_1^n) \in \aep} \P\bigl\{X^n_1(1)=x_1^n, Y^n_1(j)=y^n_1 | \Mc \bigr\} 2^{-n (I(X_2,U_3;Y_1,X_1) + I(X_2;U_3) -\d( \epsilon) ) }  \nn \\
&  \le (1+\e) 2^{n ( R_2 + R_3 -  I(X_2,U_3;Y_1,X_1) - I(X_2;U_3) + \d( \epsilon) )  }.  \nn
 \end{align}
Here, step (a) is justified as follows.
%%%%%%%%%%%%%%%%%%%%%%%%%%%%%%%%%%%%%%%%%%
%%%%%%%%%%%%%%%%%%%%%%%%%%%%%%%%%%%%%%%%%%
Let $\Ut^n = ( U_3^n(1), Y^n_3(j-1)  )$ and $\ut^n = (\ut_3^n,\yt^n_3)$ in short.
 Then, by the law of total probability, for $l \ne 1$ and $m \ne 1$,
\begin{align}
\P& \bigl\{ X^n_1(1)=x_1^n, U^n_3(l)=u^n_3,X^n_2(m)=x^n_2, Y^n_1(j)=y^n_1 |\Mc \bigr\} \nn \\
& =  \sum_{ \ut^n }  \P \bigl\{ X^n_1(1)=x_1^n, U^n_3(l)=u^n_3,X^n_2(m)=x^n_2, Y^n_1(j)=y^n_1, \Ut^n=\ut^n  | \Mc \bigr\} \nn \\
&\stackrel{(b)}{=} \sum_{ \ut^n }  \P \bigl\{ U^n_3(l)=u^n_3| \Mc , X^n_1(1)=x_1^n ,X^n_2(m)=x^n_2, \Ut^n=\ut^n  \bigr\} \nn \\
& \quad \qquad \times \P \bigl\{ \Ut^n=\ut^n  | \Mc, X^n_1(1)=x_1^n,X^n_2(m)=x^n_2, Y^n_1(j)=y^n_1 \bigr\}  \nn \\
& \quad \qquad \times  \P \bigl\{ X^n_1(1)=x_1^n,X^n_2(m)=x^n_2, Y^n_1(j)=y^n_1 | \Mc \bigr\} \nn \\
&\stackrel{(c)}{=} \sum_{ \ut^n }  \P \bigl\{ U^n_3(l)=u^n_3| L_{3,(j-1)}=1 ,  U_3^n(1) = \ut_3^n, Y^n_3(j-1)  =\yt^n_3  \bigr\} \nn \\
& \quad \qquad \times \P \bigl\{ \Ut^n=\ut^n  | \Mc, X^n_1(1)=x_1^n,X^n_2(m)=x^n_2, Y^n_1(j)=y^n_1 \bigr\}  \nn \\
& \quad \qquad \times  \P \bigl\{ X^n_1(1)=x_1^n,X^n_2(m)=x^n_2, Y^n_1(j)=y^n_1 | \Mc \bigr\} \nn \\
&\stackrel{(d)}{\le} (1+ \e) \sum_{ \ut^n } \biggl(\prod_{i=1}^np_{U_3}( u_{3i})  \biggr)  \P \bigl\{ \Ut^n=\ut^n  | \Mc, X^n_1(1)=x_1^n,X^n_2(m)=x^n_2, Y^n_1(j)=y^n_1 \bigr\}  \nn \\
& \quad \qquad \times  \P \bigl\{ X^n_1(1)=x_1^n,X^n_2(m)=x^n_2, Y^n_1(j)=y^n_1 | \Mc \bigr\} \nn \\
&=  (1 + \e)   \biggl(\prod_{i=1}^np_{U_1}( u_{1i}) \biggr) \P \bigl\{ X^n_1(1)=x_1^n,X^n_2(m)=x^n_2, Y^n_1(j)=y^n_1 | \Mc \bigr\}    \nn
\end{align}
for $n$ sufficiently large, where $(b)$ follows from the fact that given $\Mc$
 \[
U^n_3(l) \to ( X^n_1(1),X^n_2(m) , U_3^n(L_{3,(j-1)}), Y^n_3(j-1)  )  \to Y^n_1(j)
\]
form a Markov chain for all $l_2 \neq 1$ and $m \neq 1$, step (c) follows by the independence of the sequences and the encoding procedure, while $(d)$ follows from Lemma~\ref{lem:1}.
 It follows that $\P(\Ec_4)$ tends to zero as $n\to\infty$ if
\begin{equation*}
R_2 + R_3  < I(X_2,U_3; X_1,Y_1) + I(X_2;U_3) -\d(\e).
\end{equation*}

%%%%%%%%%%%%%%%%%%%%%%%%%%%%%%
%%%%%%%%%%%%%%%%%%%%%%%%%%%%%%%%%%%%%%%%%%

By similar steps, the decoding error probability at node 2 goes to zero as $n\to\infty$ if  $R_1 <  I(X_1,U_3; Y_2|X_2) -\d(\e)$  and
\begin{align*}
R_1 + R_3 & <  I(X_1,U_3; X_2,Y_2) + I(X_1;U_3) -\d(\e).
 \end{align*}
Finally, by eliminating $R_3$ from the above inequalities, the probability of error tends to zero as $n\to\infty$ if the conditions in Theorem~\ref{thm:dm-twrc} are satisfied.

\section{Proof of Theorem~\ref{thm:diamond}}
 \label{app:diamond}
The achievability proof of Theorem~\ref{thm:diamond} uses $b$ transmissions blocks, each consisting of $n$ transmissions, as in the proof of the multihop lower bound for the relay channel~\cite{El-Gamal--Kim2011}. A sequence of $(b-1)$ messages $m_j\in[1:2^{nR}],$ $j\in[1:b-1]$, each selected independently and uniformly over $[1:2^{nR}]$ is sent over $b$ blocks. Note that the average rate over the $b$ blocks is $\frac{(b-1)}{b}R$, which can be made arbitrarily close to $R$ by letting $b\to\infty$.

{\em Codebook Generation}: Let $\e > \e' > 0$. Fix $p(x_1)p(u_2|y_2)p(u_3|y_3)$ and two encoding functions $x_2(u_2,y_2)$ and $x_3(u_3,y_3)$. We randomly and independently generate a codebook for each block. For $j\in[1:b]$, randomly and independently generate $2^{nR}$ sequences $x_1^n(m_j)$, $m_j\in[1:2^{nR}],$ each according to $\prod_{i=1}^n p_{X_1}(x_{1i})$. For each $k=1,2,$ randomly and independently generate $2^{nR_k}$ sequences $u^n_k(l_{kj})$, $l_{kj}\in[1:2^{nR_k}]$, each according to $\prod_{i=1}^n p_{U_k}(u_{ki})$.
 \medskip

{\em Encoding}: Let $m_j\in[1:2^{nR}]$ be the independent message to be sent in block $j\in[1:b-1]$. Then, the source node transmits $x^n_{1}(m_j)$.
\medskip

{\em Relay Encoding}: Upon receiving $y^n_k(j)$ in block $j\in[1:b-1]$ relay node $k$, $k=1,2$, finds an index $l_{kj}$ such that $(u^n_k(l_{kj}), y^n_k(j))\in \aepvar$. If there is more than one index, it chooses one of them at random. If there is no such index, it chooses an arbitrary index at random from $[1:2^{nR_k}]$. In block $j+1$, relay $k$ then transmits $x_{ki}= x_{ki}(u_{ki}(l_{kj}), y_{ki}(j))$ for $i\in[1:n]$.
 \medskip

{\em Decoding}: Upon receiving $y^n_4(j)$, $j\in[2:b]$, the decoder finds the unique message $\mh_{j-1}$ such that
\[
    (x^n_1(\mh_{j-1}), u^n_2(l_{2,(j-1)}), u^n_3(l_{3,(j-1)}), y^n_4(j))\in \aep,
 \]
for some $l_{2,(j-1)}\in[1:2^{nR_2}]$ and $l_{3,(j-1)}\in[1:2^{nR_3}]$.
\medskip

{\em Analysis of the probability of error}: We analyze the probability of decoding error for the message $M_{j-1}$ in block $j$, $j=2,\ldots,b$, averaged over the random codebooks and index assignments. Let $L_{2,(j-1)}$ and $L_{3,(j-1)}$ be the random variables denoting the indexes chosen in block $j-1$ at relay 2 and 3, respectively.  The decoder makes an error only if one or more of the following events occur:
 %The analysis is the same for each transmission block, so we drop the block subscript to simplify notation.  To bound the probability of error, assume without loss of generality that $M=1$.

\begin{align*}
\Ec_1&=\{(Y^n_2(j-1), U^n_2(l_2 ))\not\in \aepvar \text{ for all } l_{2}\},\\
 \Ec_2&=\{(Y^n_3(j-1), U^n_3(l_3))\not\in \aepvar \text{ for all } l_3\},\\
\Ec_3&=\{(X^n_1(M_{j-1}), U^n_2(L_{2,(j-1)}),U^n_3(L_{3,(j-1)}), Y^n_4(j))\not\in \aep\},\\
\Ec_4&=\{(X^n_1(m), U^n_2(L_{2,(j-1)}),U^n_3(L_{3,(j-1)}), Y^n_4(j)) \in \aep \text{ for some } m\neq M_{j-1}\},\\
 \Ec_5&=\{(X^n_1(m), U^n_2(l_2),U^n_3(L_{3,(j-1)}), Y^n_4(j)) \in \aep \text{ for some } l_2\neq L_{2,(j-1)}, m\neq M_{j-1}\},\\
\Ec_6&=\{(X^n_1(m), U^n_2(L_{2,(j-1)}),U^n_3(l_3), Y^n_4(j)) \in \aep \text{ for some } l_3\neq L_{3,(j-1)}, m\neq M_{j-1}\},\\
 \Ec_7&=\{(X^n_1(m), U^n_2(l_2),U^n_3(l_3), Y^n_4(j) ) \in \aep \text{ for some } l_2\neq L_{2,(j-1)}, l_{3}\neq L_{3,(j-1)}, \\
& \quad \quad  m\neq M_{j-1}\}.
\end{align*}

Then by the union of events bound, the probability of decoding error is upper bounded as
 \begin{align*}
\P( \Mh_{j-1} \neq M_{j-1}) \le \P(\Ec_1)+\P(\Ec_2)+\P(\Ec_3\cap \Ec_1^c\cap \Ec_2^c)+\P(\Ec_4)+\P(\Ec_5)+\P(\Ec_6)+\P(\Ec_7).
\end{align*}

By the covering lemma, $\P(\Ec_1)$ and $\P(\Ec_2)$ tend to zero as $n\to \infty$, if
 \begin{align*}
R_2 &> I(U_2; Y_2)+\d(\e'),\\
R_3 &> I(U_3; Y_3)+\d(\e'),
\end{align*}
respectively. By the Markov lemma, $\P(\Ec_3\cap \Ec_1^c\cap \Ec_2^c)$ tends to zero as $n\to\infty$.

By the symmetry of the random codebooks generation and random index assignment at the relays, it suffices to consider the conditional probabilities of the remaining error events conditioned on the event that
\begin{align}
 \Mc = \{ M_{j-1} = 1, L_{2,(j-1)} = 1, L_{3,(j-1)} = 1\}.
\end{align}
Then, by the packing lemma, $\P(\Ec_4)$ tends to zero as $n\to \infty$ if
\begin{equation*}
R < I(X_1; U_2, U_3, Y_4)-\d(\e).
\end{equation*}
 Next, we next bound $\P(\Ec_5)$.
%%%%%%%%%%%%%%%%%%%%%%%%%%%%%%
%%%%%%%%%%%%%%%%%%%%%%%%%%%%%%
Let $\Ut^n = ( U_2^n(1), Y^n_2(j-1), Y^n_3(j-1)  )$ and $\ut^n = (\ut_2^n,\yt^n_2,\yt^n_3 )$ in short.
Then, by the law of total probability, for $l \ne 1$ and $m \ne 1$,
 \begin{align}
\P&\bigl\{X^n_1(m)=x_1^n, U^n_2(l_2)=u^n_2,U^n_3(1)=u^n_3, Y^n_4(j)=y^n_4| \Mc \bigr\} \nonumber \\
&= \sum_{ \ut^n }   \P\bigl\{ X^n_1(m)=x_1^n, U^n_2(l_2)=u^n_2, U^n_3(1)=u^n_3, Y^n_4(j)=y^n_4, \Ut^n = \ut^n  | \Mc \bigr\} \nonumber \\
&=  \sum_{ \ut^n }  \P\bigl\{  U^n_2(l_2)=u^n_2  | \Mc, X^n_1(m)=x_1^n, U^n_3(1)=u^n_3, \Ut^n = \ut^n \bigr\} \nonumber \\[-1em]
& \quad \qquad \times  \P\bigl\{ \Ut^n = \ut^n  | \Mc , X^n_1(m)=x_1^n, U^n_3(1)=u^n_3, Y^n_4(j)=y^n_4 \bigr\} \nonumber \\[-0.5em]
& \quad \qquad \times \P\bigl\{ X^n_1(m)=x_1^n, U^n_3(1)=u^n_3, Y^n_4(j)=y^n_4  | \Mc \bigr\} \nn \\
&=  \sum_{ \ut^n }  \P\bigl\{  U^n_2(l_2)=u^n_2  | L_{2,(j-1)} = 1,  U_2^n(1), Y^n_2(j-1) \bigr\} \nonumber \\[-1em]
& \quad \qquad \times  \P\bigl\{ \Ut^n = \ut^n  | \Mc , X^n_1(m)=x_1^n, U^n_3(1)=u^n_3, Y^n_4(j)=y^n_4 \bigr\} \nonumber \\[-0.5em]
& \quad \qquad \times \P\bigl\{ X^n_1(m)=x_1^n, U^n_3(1)=u^n_3, Y^n_4(j)=y^n_4  | \Mc \bigr\} \nn \\
 %
% kirti
&\le (1+\e)  \biggl(\prod_{i=1}^np_{U_2}( u_{2i}) \biggr)  \P\bigl\{X^n_1(m)=x_1^n, U^n_3(1)=u^n_3, Y^n_4(j)=y^n_4 | \Mc \bigr\} \nn
\end{align}
for $n$ sufficiently large.  Here, $(a)$ follows from the fact that given $\Mc$
 \[
U^n_2(l_2) \to \bigl(  U^n_2(L_{2,(j-1)}),U^n_3(L_{3,(j-1)}), Y^n_2(j-1), Y^n_3(j-1) \bigr) \to Y^n_4(j)
\]
form a Markov chain for all $l_2 \ne L_{2,(j-1)}$, $(b)$ follows by the independence of the sequences
 and the encoding procedure, and $(c)$ follows by Lemma~\ref{lem:1}.
%%%%%%%%%%%%%%%%%%%%%%%%%%%%%%
%%%%%%%%%%%%%%%%%%%%%%%%%%%%%%
It follows that, for $n$ large enough,
\begin{align}
\P (\Ec_5) & = \P\bigl\{(X^n_1(m), U^n_2(l_2),U^n_3(1), Y^n_4(j)) \in \aep \text{ for some } l_{2}\neq 1, m\neq 1 | \Mc \bigr\} \nonumber \\
&\le \sum_{m=2}^{2^{nR}}\sum_{l_2=2}^{2^{nR_2}}  \sum_{(x_1^n,u^n_2,u^n_3,y^n_4) \in \aep} \P\bigl\{X^n_1(m)=x_1^n, U^n_2(l_2)=u^n_2,U^n_3(1)=u^n_3, Y^n_4(j)=y^n_4| \Mc \bigr\} \nonumber \\
&\le (1+\e) \sum_{m=2}^{2^{nR}}\sum_{l_2=2}^{2^{nR_2}}  \sum_{(x_1^n,u^n_2,u^n_3,y^n_4) \in \aep} p_{U^n_2}(u^n_2) \cdot \P\bigl\{X^n_1(m)=x_1^n, U^n_3(1)=u^n_3, Y^n_4(j)=y^n_4 | \Mc \bigr\}  \label{eq:dia_in2} \nn \\
 &= (1+\e) \sum_{m=2}^{2^{nR}}\sum_{l_2=2}^{2^{nR_2}}  \sum_{(x_1^n,u^n_2,u^n_3,y^n_4) \in \aep} p_{U^n_2}(u^n_2) \cdot p_{X_1^n}(x_1^n)  \P\bigl\{U^n_3(1)=u^n_3, Y^n_4(j)=y^n_4 | \Mc \bigr\} , \nonumber \\
&= (1+\e) 2^{n(R+R_2) }  \sum_{( u^n_2, y^n_4) \in \aep} \P\bigl\{U^n_3(1)=u^n_3, Y^n_4(j)=y^n_4 | \Mc \bigr\}  2^{-n (  I(X_1, U_2; U_3, Y_4)+I(X_1;U_2)-\d(\e)  )}, \nonumber
 \end{align}
which implies that $\P(\Ec_5)$ tends to zero as $n\to\infty$ if
\begin{equation*}
R + R_2 < I(X_1, U_2; U_3, Y_4)+I(X_1;U_2)-\d(\e).
\end{equation*}
$\P(\Ec_6)$ can be bounded in a similar manner, provided that the subscripts 1 and 2 are
 interchanged in the upper bound for $\P(\Ec_5)$. It follows that $\P(\Ec_6)$ tends to zero as $n \to \infty$ if
\begin{equation*}
R + R_3 < I(X_1, U_3; U_2, Y_4)+I(X_1;U_3)-\d(\e)
\end{equation*}
%%%%%%%%%%%%%%%%%%%%%%%%%%%%%%
 %%%%%%%%%%%%%%%%%%%%%%%%%%%%%%
Next, we bound $\P(\Ec_7)$. Let $\Ut^n = ( U_2^n(1), U_3^n(1), Y^n_2(j-1), Y^n_3(j-1)  )$ and $\ut^n = (\ut_2^n,\ut_3^n,\yt^n_2,\yt^n_3 )$ in short.
Then, by the law of total probability, for $l \ne 1$ and $m \ne 1$,
 \begin{align}
\P&\bigl\{X^n_1(m)=x_1^n, U^n_2(l_2)=u^n_2,U^n_3(l_3)=u^n_3, Y^n_4(j)=y^n_4| \Mc \bigr\} \nonumber \\
&= \sum_{ \ut^n }   \P\bigl\{ X^n_1(m)=x_1^n, U^n_2(l_2)=u^n_2, U^n_3(l_3)=u^n_3, Y^n_4(j)=y^n_4, \Ut^n = \ut^n  | \Mc \bigr\} \nonumber \\
&=  \sum_{ \ut^n }  \P\bigl\{  U^n_2(l_2)=u^n_2, U^n_3(l_3)=u^n_3  | \Mc, X^n_1(m)=x_1^n, \Ut^n = \ut^n \bigr\} \nonumber \\[-1em]
& \quad \qquad \times  \P\bigl\{ \Ut^n = \ut^n  | \Mc , X^n_1(m)=x_1^n, Y^n_4(j)=y^n_4 \bigr\} \nonumber \\[-0.5em]
& \quad \qquad \times \P\bigl\{ X^n_1(m)=x_1^n, Y^n_4(j)=y^n_4  | \Mc \bigr\} \nn \\
&=  \sum_{ \ut^n }  \P\bigl\{  U^n_2(l_2)=u^n_2  | L_{2,(j-1)} = 1,  U_2^n(1), Y^n_2(j-1) \bigr\} \nonumber \\[-1em]
& \quad \qquad \times  \P\bigl\{  U^n_3(l_3)=u^n_3  | L_{3,(j-1)} = 1,  U_3^n(1), Y^n_3(j-1) \bigr\} \nonumber \\[-0.5em]
& \quad \qquad \times  \P\bigl\{ \Ut^n = \ut^n  | \Mc , X^n_1(m)=x_1^n, Y^n_4(j)=y^n_4 \bigr\} \nonumber \\[-0.5em]
& \quad \qquad \times \P\bigl\{ X^n_1(m)=x_1^n, Y^n_4(j)=y^n_4  | \Mc \bigr\} \nn \\
&\le (1+\e)  \biggl(\prod_{i=1}^np_{U_2}( u_{2i}) p_{U_3}( u_{3i}) \biggr)  \P\bigl\{X^n_1(m)=x_1^n, Y^n_4(j)=y^n_4 | \Mc \bigr\} \nn
 \end{align}
for $n$ sufficiently large.  given $\Mc$
\[
\bigl( U^n_2(l_2), U^n_2(l_3) \bigr) \to \bigl(  U^n_2(L_{2,(j-1)}),U^n_3(L_{3,(j-1)}), Y^n_2(j-1), Y^n_3(j-1) \bigr) \to Y^n_4(j)
\]
form a Markov chain for all $l_2 \ne L_{2,(j-1)}$ and  $l_3 \ne L_{3,(j-1)}$, $(b)$ follows by the independence of the sequences
 and the encoding procedure, and $(c)$ follows by applying Lemma~\ref{lem:1} twice.
It follows that, for $n$ large enough,
\begin{align}
\P (\Ec_5) & = \P\bigl\{(X^n_1(m), U^n_2(l_2),U^n_3(1), Y^n_4(j)) \in \aep \text{ for some } l_{2}\neq 1, l_{3}\neq 1, m\neq 1 | \Mc \bigr\} \nonumber \\
&\le \sum_{m=2}^{2^{nR}}\sum_{l_2=2}^{2^{nR_2}} \sum_{l_2=2}^{2^{nR_3}}  \sum_{(x_1^n,u^n_2,u^n_3,y^n_4) \in \aep} \P\bigl\{X^n_1(m)=x_1^n, U^n_2(l_2)=u^n_2,U^n_3(l_3)=u^n_3, Y^n_4(j)=y^n_4| \Mc \bigr\} \nonumber \\
&\le (1+\e) \sum_{m=2}^{2^{nR}}\sum_{l_2=2}^{2^{nR_2}}  \sum_{l_2=2}^{2^{nR_3}}  \sum_{(x_1^n,u^n_2,u^n_3,y^n_4) \in \aep} p_{U^n_2}(u^n_2) p_{U^n_3}(u^n_3) \cdot \P\bigl\{X^n_1(m)=x_1^n, Y^n_4(j)=y^n_4 | \Mc \bigr\}   \nn \\
&\le (1+\e) \sum_{m=2}^{2^{nR}}\sum_{l_2=2}^{2^{nR_2}}  \sum_{l_2=2}^{2^{nR_3}}  \sum_{(x_1^n,u^n_2,u^n_3,y^n_4) \in \aep} p_{U^n_2}(u^n_2) p_{U^n_3}(u^n_3) p_{X_1^n}(x_1^n)  \cdot \P\bigl\{ Y^n_4(j)=y^n_4 | \Mc \bigr\}   \nn \\
&= (1+\e) 2^{n(R + R_2 + R_3) }  \sum_{y^n_4 \in \aep} \P\bigl\{Y^n_4(j)=y^n_4 | \Mc \bigr\}  2^{-n (  I(X_1, U_2, U_3; Y_4)+I(X_1;U_2)+I(X_1,U_2;U_3) -\d(\e)  )}, \nonumber
\end{align}
which implies that $\P(\Ec_5)$ tends to zero as $n\to\infty$ if
 \begin{equation*}
R + R_2 + R_3< I(X_1, U_2, U_3; Y_4)+I(X_1;U_2)+I(X_1,U_2;U_3)-\d(\e),
\end{equation*}
%%%%%%%%%%%%%%%%%%%%%%%%%%%%%%
%%%%%%%%%%%%%%%%%%%%%%%%%%%%%%
 Finally, by eliminating $R_2$ and $R_3$, the probability of error tends to zero as $n\to\infty$ if the conditions in Theorem~\ref{thm:diamond} are satisfied.

%By similar steps, $\P(\Ec_6)$ and $\P(\Ec_7)$ tend to zero as $n\to\infty$ if
%\begin{equation*}
%R + R_3 < I(X_1, U_3; U_2, Y_4)+I(X_1;U_3)-\d(\e)
%\end{equation*}
%and
%\begin{equation*}
%R + R_2 + R_3< I(X_1, U_2, U_3; Y_4)+I(X_1;U_2)+I(X_1,U_2;U_3)-\d(\e),
 %\end{equation*}
%respectively. Finally, by eliminating $R_2$ and $R_3$, the probability of error tends to zero as $n\to\infty$ if the conditions in Theorem~\ref{thm:diamond} are satisfied.

%%%%%%%%%%%%%%%%%%

\bibliographystyle{IEEEtran}
\bibliography{bib_cs}

% Generated by IEEEtran.bst, version: 1.13 (2008/09/30)
\begin{thebibliography}{10}
\providecommand{\url}[1]{#1}
\csname url@samestyle\endcsname
\providecommand{\newblock}{\relax}
\providecommand{\bibinfo}[2]{#2}
\providecommand{\BIBentrySTDinterwordspacing}{\spaceskip=0pt\relax}
\providecommand{\BIBentryALTinterwordstretchfactor}{4}
\providecommand{\BIBentryALTinterwordspacing}{\spaceskip=\fontdimen2\font plus
\BIBentryALTinterwordstretchfactor\fontdimen3\font minus
  \fontdimen4\font\relax}
\providecommand{\BIBforeignlanguage}[2]{{%
\expandafter\ifx\csname l@#1\endcsname\relax
\typeout{** WARNING: IEEEtran.bst: No hyphenation pattern has been}%
\typeout{** loaded for the language `#1'. Using the pattern for}%
\typeout{** the default language instead.}%
\else
\language=\csname l@#1\endcsname
\fi
#2}}
\providecommand{\BIBdecl}{\relax}
\BIBdecl

\bibitem{Shannon1948}
C.~E. Shannon, ``A mathematical theory of communication,'' \emph{Bell System
  Tech. J.}, vol.~27, pp. 379--423, 623--656, 1948.

\bibitem{Shannon1959}
------, ``Coding theorems for a discrete source with a fidelity criterion,'' in
  \emph{IRE Int. Conv. Rec., part 4}, 1959, vol.~7, pp. 142--163, reprinted
  with changes in {\em Information and Decision Processes,} R. E. Machol, Ed.
  New York: McGraw-Hill, 1960, pp. 93-126.

\bibitem{Gastpar2003}
M.~Gastpar, B.~Rimoldi, and M.~Vetterli, ``To code, or not to code: lossy
  source-channel communication revisited,'' \emph{{IEEE} Trans. Inf. Theory},
  vol.~49, no.~5, pp. 1147--1158, May 2003.

\bibitem{Song--Yeung--Cai2006}
L.~Song, R.~Yeung, and N.~Cai, ``A separation theorem for single-source network
  coding,'' \emph{{IEEE} Trans. Inf. Theory}, vol.~52, no.~5, pp. 1861--1871,
  May 2006.

\bibitem{Ramamoorthy2006}
A.~Ramamoorthy, K.~Jain, P.~Chou, and M.~Effros, ``Separating distributed
  source coding from network coding,'' \emph{{IEEE} Trans. Inf. Theory},
  vol.~52, no.~6, pp. 2785--2795, June 2006.

\bibitem{agarwal--mitter2010}
\BIBentryALTinterwordspacing
M.~Agarwal and S.~K. Mitter, ``Communication to within a fidelity criterion
  over unknown networks by reduction to reliable communication problems over
  unknown networks,'' 2010. [Online]. Available:
  \url{http://arxiv.org/abs/1002.1300}
\BIBentrySTDinterwordspacing

\bibitem{Jalali--Effros2010}
S.~Jalali and M.~Effros, ``On the separation of lossy source-network coding and
  channel coding in wireline networks,'' in \emph{Proc. IEEE International
  Symposium on Information Theory}, June 2010, pp. 500--504.

\bibitem{Tian--Diggavi--Shamai2010}
\BIBentryALTinterwordspacing
C.~Tian, J.~Chen, S.~Diggavi, and S.~Shamai, ``Optimality and approximate
  optimality of source--channel separation in networks,'' 2010, submitted to
  {\em IEEE Trans. Inf. Theory,} 2010. [Online]. Available:
  \url{http://arxiv.org/abs/1004.2648}
\BIBentrySTDinterwordspacing

\bibitem{Cover--El-Gamal--Salehi1980}
T.~M. Cover, A.~El~Gamal, and M.~Salehi, ``Multiple access channels with
  arbitrarily correlated sources,'' \emph{{IEEE} Trans. Inf. Theory}, vol.~26,
  no.~6, pp. 648--657, Nov. 1980.

\bibitem{deBruyn--prelov1987}
K.~de~Bruyn, V.~Prelov, and E.~van~der Meulen, ``Reliable transmission of two
  correlated sources over an asymmetric multiple-access channel (corresp.),''
  \emph{{IEEE} Trans. Inf. Theory}, vol.~33, no.~5, pp. 716 -- 718, Sep. 1987.

\bibitem{Rajesh--etal2008}
R.~Rajesh, V.~Varshneya, and V.~Sharma, ``Distributed joint source channel
  coding on a multiple access channel with side information,'' in \emph{Proc.
  IEEE International Symposium on Information Theory}, Toronto, Canada, July
  2008, pp. 2707--2711.

\bibitem{RajeshSharma2009}
\BIBentryALTinterwordspacing
R.~Rajesh, V.~Sharma, and V.~K. Varshenya, ``Joint source-channel coding on a
  multiple access channel with side information,'' 2009, submitted to {\em IEEE
  Trans. Inf. Theory,} 2007. [Online]. Available:
  \url{http://arxiv.org/pdf/0904.4006/}
\BIBentrySTDinterwordspacing

\bibitem{Lapidoth--Tinguely2010}
A.~Lapidoth and S.~Tinguely, ``Sending a bivariate {G}aussian over a {G}aussian
  {MAC},'' \emph{{IEEE} Trans. Inf. Theory}, vol.~56, no.~6, pp. 2714--2752,
  2010.

\bibitem{Lapidoth--Tinguely2010b}
------, ``Sending a bivariate {G}aussian source over a {G}aussian {MAC} with
  feedback,'' \emph{{IEEE} Trans. Inf. Theory}, vol.~56, no.~4, pp. 1852--1864,
  2010.

\bibitem{Lim--Minero--Kim2010}
S.~H. Lim, P.~Minero, and Y.-H. Kim, ``Lossy communication of correlated
  sources over multiple access channels,'' in \emph{Proc. 48th Annual Allerton
  Conference on Communications, Control, and Computing}, Monticello, IL, Oct.
  2010.

\bibitem{Jain-etal2012}
A.~Jain, D.~G{\"u}nd{\"u}z, S.~Kulkarni, H.~Poor, and S.~Verd{\'u},
  ``Energy-distortion tradeoffs in {G}aussian joint source-channel coding
  problems,'' \emph{{IEEE} Trans. Inf. Theory}, vol.~58, no.~5, pp. 3153--3168,
  May 2012.

\bibitem{Han--Costa1987}
T.~S. Han and M.~M.~H. Costa, ``Broadcast channels with arbitrarily correlated
  sources,'' \emph{{IEEE} Trans. Inf. Theory}, vol.~33, no.~5, pp. 641--650,
  Sep. 1987.

\bibitem{Mittal--Phamdo2002}
U.~Mittal and N.~Phamdo, ``Hybrid digital-analog ({HDA}) joint source-channel
  codes for broadcasting and robust communications,'' \emph{{IEEE} Trans. Inf.
  Theory}, vol.~48, no.~5, pp. 1082--1102, May 2002.

\bibitem{Tuncel2006}
E.~Tuncel, ``Slepian-{W}olf coding over broadcast channels,'' \emph{{IEEE}
  Trans. Inf. Theory}, vol.~52, no.~4, pp. 1469--1482, Apr. 2006.

\bibitem{Wei--Kramer2008}
K.~Wei and G.~Kramer, ``Broadcast channel with degraded source random variables
  and receiver side information,'' in \emph{Proc. IEEE International Symposium
  on Information Theory}, Toronto, Canada, July 2008, pp. 1711--1715.

\bibitem{Kramer--Nair2009}
G.~Kramer and C.~Nair, ``Comments on ``{B}roadcast channels with arbitrarily
  correlated sources'','' in \emph{Proc. IEEE International Symposium on
  Information Theory}, Seoul, Korea, July 2009, pp. 2777--2779.

\bibitem{Minero--Kim2009}
P.~Minero and Y.-H. Kim, ``Correlated sources over broadcast channels,'' in
  \emph{Proc. IEEE International Symposium on Information Theory}, Seoul,
  Korea, July 2009, pp. 2780--2784.

\bibitem{Kramer--Liang--Shamai2009}
G.~Kramer, Y.~Liang, and S.~Shamai, ``Outer bounds on the admissible source
  region for broadcast channels with dependent sources,'' in \emph{Information
  Theory and Applications Workshop, 2009}, Feb. 2009, pp. 169--172.

\bibitem{Soundararajan-Vishwanath2009}
R.~Soundararajan and S.~Vishwanath, ``Hybrid coding for {G}aussian broadcast
  channels with {G}aussian sources,'' in \emph{Proc. IEEE International
  Symposium on Information Theory}, Seoul, Korea, July 2009, pp. 2790 --2794.

\bibitem{Tian--Diggavi--Shamai2010b}
C.~Tian, S.~Diggavi, and S.~Shamai, ``The achievable distortion region of
  bivariate {G}aussian source on {G}aussian broadcast channel,'' \emph{{IEEE}
  Trans. Inf. Theory}, vol.~57, no.~10, pp. 6419--6427, Oct. 2011.

\bibitem{Nayak--Tuncel--Gunduz2010}
J.~Nayak, E.~Tuncel, and D.~G{\"u}nd{\"u}z, ``Wyner-{Z}iv coding over broadcast
  channels: Digital schemes,'' \emph{{IEEE} Trans. Inf. Theory}, vol.~56,
  no.~4, pp. 1782--1799, Apr. 2010.

\bibitem{Gao--Turcel2011b}
Y.~Gao and E.~Tuncel, ``{W}yner-{Z}iv coding over broadcast channels: hybrid
  digital/analog schemes,'' \emph{{IEEE} Trans. Inf. Theory}, vol.~57, no.~9,
  pp. 5660--5672, Sept. 2011.

\bibitem{Liu--Chen2010}
W.~Liu and B.~Chen, ``Communicating correlated sources over interference
  channels: The lossy case,'' in \emph{Proc. IEEE International Symposium on
  Information Theory}, Austin, Texas, June 2010, pp. 345 --349.

\bibitem{Liu--Chen2011}
------, ``Interference channels with arbitrarily correlated sources,''
  \emph{{IEEE} Trans. Inf. Theory}, vol.~57, no.~12, pp. 8027--8037, Dec. 2011.

\bibitem{Gunduz--etal2009}
D.~G{\"u}nd{\"u}z, E.~Erkip, A.~Goldsmith, and H.~Poor, ``Source and channel
  coding for correlated sources over multiuser channels,'' \emph{{IEEE} Trans.
  Inf. Theory}, vol.~55, no.~9, pp. 3927--3944, Sept. 2009.

\bibitem{Coleman--elal2009}
T.~Coleman, E.~Martinian, and E.~Ordentlich, ``Joint source-channel coding for
  transmitting correlated sources over broadcast networks,'' \emph{{IEEE}
  Trans. Inf. Theory}, vol.~55, no.~8, pp. 3864--3868, Aug. 2009.

\bibitem{Gunduzetal-2013}
D.~G{\"u}nd{\"u}z, E.~Erkip, A.~Goldsmith, and H.~Poor, ``Reliable joint
  source--channel cooperative transmission over relay networks,'' \emph{{IEEE}
  Trans. Inf. Theory}, vol.~59, no.~4, pp. 2442--2458, Apr. 2013.

\bibitem{Gelfand--Pinsker1980a}
S.~I. Gelfand and M.~S. Pinsker, ``Coding for channel with random parameters,''
  \emph{Probl. Control Inf. Theory}, vol.~9, no.~1, pp. 19--31, 1980.

\bibitem{Shannon1958a}
C.~E. Shannon, ``Channels with side information at the transmitter,'' \emph{IBM
  J. Res. Develop.}, vol.~2, pp. 289--293, 1958.

\bibitem{Gray--Wyner1974}
R.~M. Gray and A.~D. Wyner, ``Source coding for a simple network,'' \emph{Bell
  System Tech. J.}, vol.~53, pp. 1681--1721, 1974.

\bibitem{Wilson--etal2010}
M.~Wilson, K.~Narayanan, and G.~Caire, ``Joint source channel coding with side
  information using hybrid digital analog codes,'' \emph{{IEEE} Trans. Inf.
  Theory}, vol.~56, no.~10, pp. 4922--4940, Oct. 2010.

\bibitem{Dueck1981a}
G.~Dueck, ``A note on the multiple access channel with correlated sources,''
  \emph{{IEEE} Trans. Inf. Theory}, vol.~27, no.~2, pp. 232--235, 1981.

\bibitem{Lim--Kim--El-Gamal--Chung2011}
S.~H. Lim, Y.-H. Kim, A.~El~Gamal, and S.-Y. Chung, ``Noisy network coding,''
  \emph{{IEEE} Trans. Inf. Theory}, vol.~57, no.~5, pp. 3132--3152, May 2011.

\bibitem{schein--gallager2000}
B.~Schein and R.~Gallager, ``The {G}aussian parallel relay network,'' in
  \emph{IEEE International Symposium on Information Theory, 2000}, 2000, p.~22.

\bibitem{Cover--El-Gamal1979}
T.~M. Cover and A.~El~Gamal, ``Capacity theorems for the relay channel,''
  \emph{{IEEE} Trans. Inf. Theory}, vol.~25, no.~5, pp. 572--584, Sep. 1979.

\bibitem{Rankov--Wittneben2006}
B.~Rankov and A.~Wittneben, ``Achievable rate regions for the two-way relay
  channel,'' in \emph{Proc. IEEE International Symposium on Information
  Theory}, Seattle, WA, Jul. 2006, pp. 1668--1672.

\bibitem{El-Gamal--Kim2011}
A.~El~Gamal and Y.-H. Kim, \emph{Network Information Theory}.\hskip 1em plus
  0.5em minus 0.4em\relax Cambridge University Press, 2011.

\bibitem{Orlitsky--Roche2001}
A.~Orlitsky and J.~R. Roche, ``Coding for computing,'' \emph{{IEEE} Trans. Inf.
  Theory}, vol.~47, no.~3, pp. 903--917, 2001.

\bibitem{Goblick1965}
T.~J. Goblick, ``Theoretical limitations on the transmission of data from
  analog sources,'' \emph{{IEEE} Trans. Inf. Theory}, vol.~11, pp. 558--567,
  1965.

\bibitem{FaT}
P.~Minero, S.~H. Lim, and Y.-H. Kim, ``Hybrid coding: An interface for joint
  source-channel coding and network communication,'' 2013, to be submitted to
  Foundations and Trends in Communications and Information Theory.

\bibitem{Wyner--Ziv1976}
A.~D. Wyner and J.~Ziv, ``The rate-distortion function for source coding with
  side information at the decoder,'' \emph{{IEEE} Trans. Inf. Theory}, vol.~22,
  no.~1, pp. 1--10, 1976.

\bibitem{Merhav--Shamai2003}
N.~Merhav and S.~Shamai, ``On joint source-channel coding for the {W}yner-{Z}iv
  source and the {G}elfand-{P}insker channel,'' \emph{{IEEE} Trans. Inf.
  Theory}, vol.~49, no.~11, pp. 2844--2855, Nov. 2003.

\bibitem{Sutivong-etal-2005}
A.~Sutivong, M.~Chiang, T.~Cover, and Y.-H. Kim, ``Channel capacity and state
  estimation for state-dependent {G}aussian channels,'' \emph{{IEEE} Trans.
  Inf. Theory}, vol.~51, no.~4, pp. 1486--1495, April 2005.

\bibitem{Berger1978}
T.~Berger, ``Multiterminal source coding,'' in \emph{The Information Theory
  Approach to Communications}, G.~Longo, Ed.\hskip 1em plus 0.5em minus
  0.4em\relax New York: Springer-Verlag, 1978.

\bibitem{Tung1978}
S.-Y. Tung, ``Multiterminal source coding,'' Ph.D. Thesis, Cornell University,
  Ithaca, NY, 1978.

\bibitem{Gacs--Korner1973}
P.~G{\'a}cs and J.~K{\"o}rner, ``Common information is far less than mutual
  information,'' \emph{Probl. Control Inf. Theory}, vol.~2, no.~2, pp.
  149--162, 1973.

\bibitem{Witsenhausen1975}
H.~S. Witsenhausen, ``On sequences of pairs of dependent random variables,''
  \emph{SIAM J. Appl. Math.}, vol.~28, pp. 100--113, 1975.

\bibitem{Wagner--Kelly--Altug2011}
A.~B. Wagner, B.~G. Kelly, and Y.~Altu\u{g}, ``Distributed rate-distortion with
  common components,'' \emph{{IEEE} Trans. Inf. Theory}, vol.~57, no.~7, pp.
  4035--4057, Aug. 2011.

\bibitem{Slepian--Wolf1973b}
D.~Slepian and J.~K. Wolf, ``A coding theorem for multiple access channels with
  correlated sources,'' \emph{Bell System Tech. J.}, vol.~52, pp. 1037--1076,
  Sep. 1973.

\bibitem{Berger--Zhang--Viswanathan1996}
T.~Berger, Z.~Zhang, and H.~Viswanathan, ``The {CEO} problem,'' \emph{{IEEE}
  Trans. Inf. Theory}, vol.~42, no.~3, pp. 887--902, 1996.

\bibitem{Viswanathan--Berger1997}
H.~Viswanathan and T.~Berger, ``The quadratic {G}aussian {CEO} problem,''
  \emph{{IEEE} Trans. Inf. Theory}, vol.~43, no.~5, pp. 1549--1559, 1997.

\bibitem{Prabhakaran--Tse--Ramchandran2004}
V.~Prabhakaran, D.~N.~C. Tse, and K.~Ramchandran, ``Rate region of the
  quadratic {G}aussian {CEO} problem,'' in \emph{Proc. IEEE International
  Symposium on Information Theory}, Chicago, IL, June/July 2004, p. 117.

\bibitem{Oohama2005}
Y.~Oohama, ``Rate-distortion theory for {G}aussian multiterminal source coding
  systems with several side informations at the decoder,'' \emph{{IEEE} Trans.
  Inf. Theory}, vol.~51, no.~7, pp. 2577--2593, Jul. 2005.

\bibitem{Gastpar2008}
M.~Gastpar, ``Uncoded transmission is exactly optimal for a simple {G}aussian
  ``sensor'' network,'' \emph{Information Theory, IEEE Transactions on},
  vol.~54, no.~11, pp. 5247--5251, Nov. 2008.

\bibitem{Marton1979}
K.~Marton, ``A coding theorem for the discrete memoryless broadcast channel,''
  \emph{{IEEE} Trans. Inf. Theory}, vol.~25, no.~3, pp. 306--311, 1979.

\bibitem{Aref1980}
M.~R. Aref, ``Information flow in relay networks,'' Ph.D. Thesis, Stanford
  University, Stanford, CA, Oct. 1980.

\bibitem{Kramer--Gastpar--Gupta2005}
G.~Kramer, M.~Gastpar, and P.~Gupta, ``Cooperative strategies and capacity
  theorems for relay networks,'' \emph{{IEEE} Trans. Inf. Theory}, vol.~51,
  no.~9, pp. 3037--3063, Sep. 2005.

\bibitem{Nazer--Gastpar2007}
B.~Nazer and M.~Gastpar, ``Computation over multiple-access channels,''
  \emph{{IEEE} Trans. Inf. Theory}, vol.~53, no.~10, pp. 3498--3516, Oct. 2007.

\bibitem{Nam--Chung--Lee2009}
W.~Nam, S.-Y. Chung, and Y.~H. Lee, ``Capacity of the {G}aussian two-way relay
  channel within 1/2 bit,'' \emph{{IEEE} Trans. Inf. Theory}, vol.~56, no.~11,
  pp. 5488--5494, Nov. 2010.

\bibitem{Laneman-etal2004}
J.~Laneman, D.~Tse, and G.~W. Wornell, ``Cooperative diversity in wireless
  networks: Efficient protocols and outage behavior,'' \emph{{IEEE} Trans. Inf.
  Theory}, vol.~50, no.~12, pp. 3062--3080, 2004.

\bibitem{Khormuji--Skoglund2010}
M.~Khormuji and M.~Skoglund, ``On instantaneous relaying,'' \emph{{IEEE} Trans.
  Inf. Theory}, vol.~56, no.~7, pp. 3378--3394, July 2010.

\bibitem{Khormuji--Skoglund2011}
------, ``Hybrid digital-analog noisy network coding,'' in \emph{2011
  International Symposium on Network Coding (NetCod)}, July 2011, pp. 1--5.

\bibitem{El-Gamal1981b}
A.~El~Gamal, ``On information flow in relay networks,'' in \emph{Proc. IEEE
  National Telecom Conference}, Nov. 1981, vol.~2, pp. D4.1.1--D4.1.4.

\bibitem{Avestimehr--Diggavi--Tse2009}
S.~Avestimehr, S.~Diggavi, and D.~Tse, ``Wireless network information flow: {A}
  deterministic approach,'' \emph{{IEEE} Trans. Inf. Theory}, vol.~57, no.~4,
  pp. 1872--1905, Apr. 2011.

\bibitem{Kaspi--Berger1982}
A.~H. Kaspi and T.~Berger, ``Rate-distortion for correlated sources with
  partially separated encoders,'' \emph{{IEEE} Trans. Inf. Theory}, vol.~28,
  no.~6, pp. 828--840, 1982.

\bibitem{Wagner--Kelly--Altug2009}
A.~B. Wagner, B.~G. Kelly, and Y.~Altu\u{g}, ``The lossy one-helper conjecture
  is false,'' in \emph{Proc. 47th Annual Allerton Conference on Communications,
  Control, and Computing}, Monticello, IL, Sep. 2009.

\bibitem{Han--Kobayashi1981}
T.~S. Han and K.~Kobayashi, ``A new achievable rate region for the interference
  channel,'' \emph{{IEEE} Trans. Inf. Theory}, vol.~27, no.~1, pp. 49--60,
  1981.

\end{thebibliography}

\end{document}